\documentclass[english,aps,prb,twocolumn]{revtex4-1}
\usepackage{mathptmx}
\usepackage{berasans}
\usepackage{beramono}

\usepackage[T1]{fontenc}
\usepackage[latin9]{inputenc}
\usepackage{listings}
\usepackage{xcolor}
\usepackage{pdfcolmk}
\usepackage{babel}
\usepackage{fancybox}
\usepackage{calc}
\usepackage{textcomp}
\usepackage{amstext}
\usepackage{amssymb}
\usepackage{graphicx}
\usepackage{setspace}
\PassOptionsToPackage{normalem}{ulem}
\usepackage{ulem}
\usepackage{subscript}
\usepackage[unicode=true,pdfusetitle,
 bookmarks=true,bookmarksnumbered=false,bookmarksopen=false,
 breaklinks=true,pdfborder={0 0 0},backref=false,colorlinks=true]
 {hyperref}

\makeatletter

\providecolor{lyxadded}{rgb}{0,0,1}
\providecolor{lyxdeleted}{rgb}{1,0,0}

\@ifundefined{textcolor}{}
{%
 \definecolor{BLACK}{gray}{0}
 \definecolor{WHITE}{gray}{1}
 \definecolor{RED}{rgb}{1,0,0}
 \definecolor{GREEN}{rgb}{0,1,0}
 \definecolor{BLUE}{rgb}{0,0,1}
 \definecolor{CYAN}{cmyk}{1,0,0,0}
 \definecolor{MAGENTA}{cmyk}{0,1,0,0}
 \definecolor{YELLOW}{cmyk}{0,0,1,0}
}

\renewcommand{\Im}{\ensuremath{\mathfrak{Im}}}
\renewcommand{\Re}{\ensuremath{\mathfrak{Re}}}
\definecolor{darkred}{rgb}{0.5,0.2,0.2}
\definecolor{red}{rgb}{0.5,0.2,0.2}

\makeatother

\begin{document}

\title{Finding the bare band: Electron coupling to two phonon modes in potassium-doped
graphene on Ir(111)}

\author{I. Pletikosi\'{c}}

\email{ivo.pletikosic@ifs.hr}

\author{M. Kralj}

\author{M. Milun}

\author{P. Pervan}

\affiliation{Institut za fiziku, Bijeni\v{c}ka 46, HR-10000 Zagreb, Croatia }
\begin{abstract}
We analyze renormalization of the $\pi^{*}$ band of \emph{n}-doped
epitaxial graphene on Ir(111) induced by electron-phonon coupling.
Our procedure of extracting the bare band relies on recursive self-consistent
refining of the functional form of the bare band until the convergence.
We demonstrate that the components of the self-energy,  as well as
the spectral intensity  obtained from angle-resolved photoelectron
spectroscopy (ARPES) show that the renormalization is due to the coupling
to two distinct phonon excitations. From the velocity renormalization
and an increase of the imaginary part of the self-energy  we find
the electron-phonon coupling constant to be $\sim0.2$, which is in
fair agreement with a previous study of the same system, despite the
notable difference in the width of spectroscopic curves. Our experimental
results also suggest that potassium intercalated between graphene
and Ir(111) does not introduce any additional increase of the quasiparticle
scattering rate. 
\end{abstract}


\maketitle

\section{Introduction}

Graphene is a fascinating two-dimensional material with linear electronic
bands and linear dependence of the density of states around the Fermi
level. \citep{CastroNeto2009} By field-effect or chemical doping
(either \emph{n} or \emph{p}) Fermi level can be tuned to change graphene
from a zero-gap semiconductor to a metal. \citep{Giovannetti2008,Coletti2010,Liu2011a}
Intrinsic and doped graphene prove to be an interesting platform to
study different aspects of many-body interactions in two-dimensional
systems, either experimentally \citep{Bostwick2007c,Bostwick2010,Siegel2011}
or theoretically \citep{DasSarma2007,Mak2011,Park2008c} . A particular
accent is put on the electron-phonon coupling (EPC) in doped graphene.
\citep{Gruneis2009,Valla2009,Borysenko2010}

Adsorption and/or intercalation of alkali atoms in epitaxial graphenes
can lead to a whole range of \emph{n} dopings, including an extreme
case where graphene\textquoteright{}s van Hove point is brought to
the Fermi level. \citep{McChesney2010} Studies on KC\textsubscript{8}
(\citet{Gruneis2009}) and CaC\textsubscript{6} (\citet{Valla2009})
report strong EPC anisotropy, with significantly smaller EPC constant
$\lambda$ along the K-$\Gamma$, compared to the K-M direction. Despite
the fact that these experiments were performed by intercalation of
graphite, it had been demonstrated that the intercalation separates
single layers of graphite in such a way that it shows all signatures
of graphene. Some others  have reported smaller value of $\lambda$,\citep{Bianchi2010,Zhou2008a,Siegel2011a}
and no anisotropy in the EPC. \citep{Bianchi2010} Recent results
also suggest that EPC progressively increases with doping. \citep{Pan2011}

Determination of the renormalization effects in \emph{n}-doped $\pi^{*}$
band due to the EPC is complicated, in particular along the K-M direction,
because of the non-linear behavior of the band close to the Fermi
level. \citep{Park2008a} A simple procedure to establish the strength
of electron-phonon coupling by a comparison of the quasiparticle velocity
at the Fermi level and the velocity well beyond the phonon energy
scale is shown to be compromised by a significant change of the quasiparticle
velocity due to the nonlinearity of the $\pi^{*}$ band. \citep{Park2008a}
Some other procedures rely on the bare-band dispersion. \citep{Hofmann2009}
The choice of the bare-band may, however, influence an accurate determination
of the magnitude and shape of the self-energy. \citep{Kordyuk2005}
A self-consistent GW approximation was applied in \emph{ab-initio}
density functional theory (DFT) calculation to model graphene's bare
band that includes all electron-electron correlations. \citep{Gruneis2009}
In order to avoid any arbitrariness, determination of the bare band
dispersion from the experimental data is desirable. Different approaches
based on self-consistent procedures have been developed to determine
the bare band, self-energy and ultimately, electron-phonon coupling
strength. \citep{Kordyuk2005,Veenstra2011,Bianchi2010} 

A self-consistent method has already been applied to the photoemission
data of potassium doped graphene on Ir(111). \citep{Bianchi2010}
Several aspects of the low-energy quasiparticle dynamics were addressed:
renormalization of the $\pi^{*}$ band close to the Fermi level due
to the coupling to phonons; phonon spectrum associated with the renormalization;
the width of spectral lines in connection with the electron scattering
rate. A model with a spectrum of five evenly spaced phonons participating
in the coupling to graphene\textquoteright{}s $\pi^{*}$ band was
used in the self-energy analysis. The EPC constant was found relatively
small (0.28) and isotropic around K. Rather broad peaks were interpreted
in terms of an increased electron scattering rate caused by the loss
of translation symmetry induced by the incommensurability of the system
graphene/Ir(111). \citep{Bianchi2010} An ARPES study on another metallic
system --- graphene on a copper foil --- questions all previous findings,
proposing an order of magnitude smaller electron-phonon coupling strength.
\citep{Siegel2011a}

In this paper we present results of an ARPES study of highly ordered
graphene on Ir(111)\citep{Kralj2011,Pletikosic2009} intercalated
\citep{Tontegode1993} with potassium. We use maxima of momentum distribution
curves (MDC) and energy distribution curves (EDC) to determine the
exact dispersion of the $\pi^{*}$ band along the K-M direction. 
Peak positions and widths of MDCs are used in a self-consistent method
to reconstruct the bare band $E_{b}$ and the corresponding $\Im\Sigma(E)$
and $\Re\Sigma(E)$. Both show, consistently with the spectral intensity
$A(E)$, that the renormalization is due to the coupling to two distinct
phonon excitations. From the velocity renormalization and an increase
of $\Im\Sigma(E)$ with energy electron-phonon coupling constant is
determined.

\section{Method}

Photoemission spectra offer a wealth of information about many-body
interactions in solids, in particular two-dimensional ones. The single-particle
spectral function, that the intensities in the photoemission spectra
are proportional to, is given by:

\begin{equation}
A(k,E)=\frac{1}{\pi}\frac{\Im\Sigma(E)}{[E-E_{b}(k)-\Re\Sigma(E)]^{2}+[\Im\Sigma(E)]^{2}}\label{eq:A}
\end{equation}
Here, $E_{b}$ is the dispersion relation of the bare  band, and
$\Sigma=\Re\Sigma+i\,\Im\Sigma$ its many-body correction, so-called
self-energy. The latter is usually taken to depend only on energy,
as its momentum dependence is considered weak. \citep{Veenstra2011} 

If a cut at a given energy $E=E_{m}$ (momentum distribution curve)
is made out of a two-dimensional map $A(k,E)$, nearly Lorentzian
lineshape is obtained with a maximum at $k_{m}$ such that

\begin{equation}
E_{m}-E_{b}(k_{m})-\Re\Sigma(E_{m})=0\label{eq:ReS}
\end{equation}
and a half-maximum at $k_{m}-w_{L\, m}$ and $k_{m}+w_{R\, m}$ ($w_{L\, m}+w_{R\, m}=2\, w_{m}$
is then full-width at half-maximum, FWHM), where

\begin{equation}
\Im\Sigma(E_{m})=E_{b}(k_{m})-E_{b}(k_{m}-w_{L\, m})\label{eq:ImS}
\end{equation}
as elaborated by Kordyuk et al.\citep{Kordyuk2005,Kordyuk2005a} Note
that these relations do not rely on any specific dispersion of the
bare band $E_{b}(k)$. 

The analysis of all the MDCs usually provides a set of value quadruplets
($E_{m}$,~$k_{m}$, $2\, w_{m}$,~ $A_{m}$) evenly spaced on $E_{m}$.
If the lineshape of an MDC is not far from Lorentzian, one has $w_{L\, m}\approx w_{R\, m}$,
and can set them both equal to $w_{m}$ --- half-width at half-maximum,
HWHM. The fact that each of the quadruplets is supposed to satisfy
equations (\ref{eq:A}), (\ref{eq:ReS}), (\ref{eq:ImS}) can help
us determine the functional form of the bare-band $E_{b}(k)$, and
the two components of the self-energy $\Re\Sigma(E)$, $\Im\Sigma(E)$
in the range of momenta and energies covered by the spectrum. 

The procedure of extracting $E_{b}$, $\Re\Sigma$, $\Im\Sigma$ from
the experimental data assumes that (i) the components of the self-energy
must, by causality, comply with the Kramers-Kronig relation over the
$\pm\infty$ range of energies; (ii) the upper half of the spectral
function not accessible to ARPES ($E>E_{\mathrm{Fermi}}\equiv0$)
is substituted by supposing particle-hole symmetry; (iii) the high-energy
tails of the three functions are not affected by a finite energy window
for which the ARPES spectrum is available. \citep{Norman1999,Kordyuk2005}

If these criteria are met, the Kramers-Kronig transform of $\Im\Sigma$
to $\Re\Sigma$ (and vice versa) 
\begin{equation}
\Re\Sigma(E)=\frac{1}{\pi}\int_{-\infty}^{+\infty}\frac{\Im\Sigma(\xi)}{\xi-E}\,\mathrm{d}\xi\label{eq:Kramers-Kronig}
\end{equation}
if considered as a convolution of two functions, is easily calculated
going through the time domain by Fourier transforms (FFT, in the discrete
case) 
\begin{equation}
\left[\Re\Sigma(E)\right]_{\mathrm{FFT}}=\left[\frac{1}{E}\right]_{\mathrm{FFT}}\,\text{\texttimes}\,\left[\Im\Sigma(E)\right]_{\mathrm{FFT}}\label{eq:FFT-KK}
\end{equation}
In some numerical packages this is already provided as a discrete
Hilbert transform.

The bare band and the components of the self-energy are usually reconstructed
by assuming a polynomial (mostly linear) form of the bare-band, from
which $\Re\Sigma(E)$ is calculated by both eq. (\ref{eq:ReS}) and
a Kramers-Kronig transform of $\Im\Sigma(E)$. Usually, not even the
eq. (\ref{eq:ImS}) is used, but its expansion to the first order
in $w$, $\Im\Sigma(E_{m})=\hbar v_{b}(E_{m})\cdot w_{m}$, where
$v_{b}(E)$ is the bare-band velocity. The two forms of $\Re\Sigma(E)$
are then compared, and the coefficients of the polynomial adjusted
until the difference is minimal. \citep{Kordyuk2005,McChesney2008,Norman1999,Hofmann2009} 

\begin{figure}
\begin{centering}
\includegraphics[width=8cm]{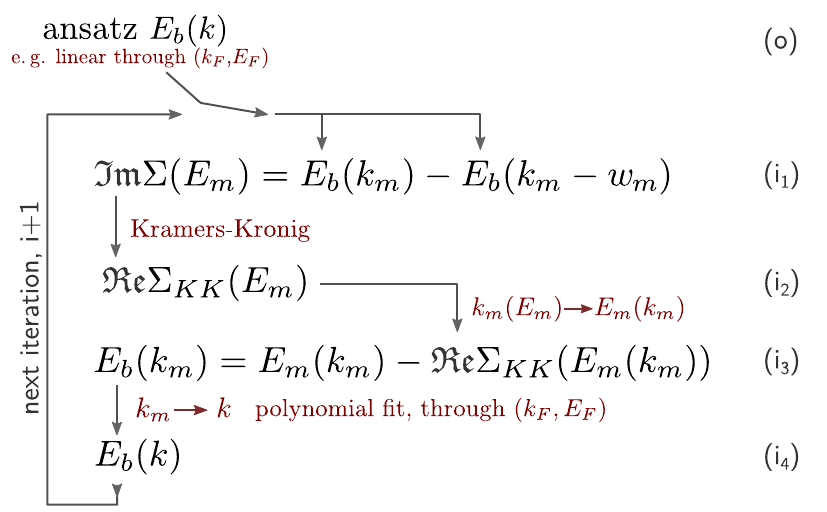}
\par\end{centering}

\caption{\label{fig:recursion-formula}(Color online) Self-consistent iterative
algorithm that refines the bare band $E_{b}$ by enforcing the Kramers-Kronig
relation between $\Im\Sigma$ and $\Re\Sigma$.}
\end{figure}
In this work we propose a simpler procedure, that avoids the need
for a minimization of a functional, and achieves the self-consistency
in just a few iterations. The procedure also appears to be more stable
with respect to the noise present in the experimental data, and no
denoising\citep{Norman1999} or smoothing by fitting to a function
\citep{Bianchi2010} is needed.

The algorithm, shown in Fig. \ref{fig:recursion-formula}, starts
by postulating a form, linear for example, of the bare band $E_{b}(k)$.
Forcing the function to pass through the experimentally determined
point ($k_{F}$,~$E_{F}$) can help in faster convergence. Note that
this constraint is physically sound, as the renormalized and the bare-band
are expected to intersect at the Fermi level. In the first step of
the iterative procedure, $\Im\Sigma$ is calculated by eq. (\ref{eq:ImS}).
In the second step $\Re\Sigma$ is obtained by a discrete Hilbert
transform of $\Im\Sigma$. This is then by eq. (\ref{eq:ReS}) used
to refine the bare band, first on a discrete set of points (step 3),
then fitted to a polynomial or a function at wish (step 4). The iteration
ends when there are no substantial changes in the functions (or parameters)
obtained. Further details on the procedure, the full code and exemplary
input data are accessible as supplemental material in Ref. \onlinecite{SupplemMat}. 

Electron-phonon coupling strength can be extracted \citep{Fink2006}
either from a steplike increase of the imaginary part of the self-energy,
$\Delta\Im\Sigma$, as 
\begin{equation}
\lambda=\frac{2}{\pi}\frac{\Delta\Im\Sigma}{\hbar\omega_{ph}}\label{eq:lambdaIm}
\end{equation}
or from the real part of the self energy, as 
\begin{equation}
\lambda=-\frac{\mathrm{d}}{\mathrm{d}E}\Re\Sigma(E)|_{E=E_{F}}\label{eq:lambdaRe}
\end{equation}
Although these two equations are strictly valid only at zero temperature,
they are especially good approximation in the case of graphene, where
phonon excitations are high in energy.

Instead of calculating the derivative in eq. (\ref{eq:lambdaRe})
directly, which is liable to errors due to the noise and scarcity
of reliable data in the very proximity of the Fermi level, one can
make use of eq. (\ref{eq:ReS}), through $\lambda=-\frac{\mathrm{d}}{\mathrm{d}E_{m}}(E_{m}-E_{b}(k_{m}))=\frac{E_{b}'(k)}{E_{m}'(k)}-1$,
to get 
\begin{equation}
\lambda=\frac{v_{b}(E_{F})}{v_{r}(E_{F})}-1\label{eq:lambdaVbVr}
\end{equation}
Here, $v_{b}$ ($v_{r}$) is the Fermi velocity of the bare (renormalized)
band.

\section{Experimental }

The experiments were performed at the APE beamline at ELETTRA. Iridium
single crystal of 99.99\% purity and surface orientation better than
0.1\textdegree{} was used. The substrate was cleaned by several cycles
of sputtering with 1.5 keV Ar\textsuperscript{+} ions at room temperature
followed by annealing at 1600 K. Cleanness and quality of Ir(111)
were checked by ARPES (existence and quality of iridium surface states)\citep{Pletikosic2010}
and low energy electron diffraction (LEED). The base pressure was
5·10\textsuperscript{\textendash{}9}~Pa.

During the growth of graphene the surface was exposed to 2 Langmuir
(L, 1L=1.3·10\textsuperscript{\textendash{}4} Pa·s) of ethene at
300 K and subsequently heated to 1470 K. In order to ensure perfectly
oriented graphene and the coverage of the whole surface, the procedure
was repeated up to 12 times, finishing with simultaneous exposure
to ethene and heating. \citep{Kralj2011,Hattab2011} Potassium was
added at room temperature by evaporation from a getter source, until
a clear 2\texttimes{}2 structure emerged in LEED and no additional
doping of graphene bands could be achieved.

ARPES spectra have been collected by a Scienta SES 2002 analyzer.
For this experiment we used photon energy of 40.5~eV, in \emph{s}
and \emph{p} polarization of the field. Typical spot size on the sample
was 50\texttimes{}100~$\mathrm{\mu m}$. The energy resolution of
this setup was about 12~meV, and the angular resolution 0.1\textdegree{}.
The azimuths were checked according to the orientation of spots in
LEED and the emergence of the Dirac cone and replica bands in ARPES
for the plain graphene on Ir(111). \citep{Pletikosic2009} During
the measurements the temperature of the sample was held at 80~K.

\section{Results}

\begin{figure}
\begin{centering}
\includegraphics[width=6.5cm]{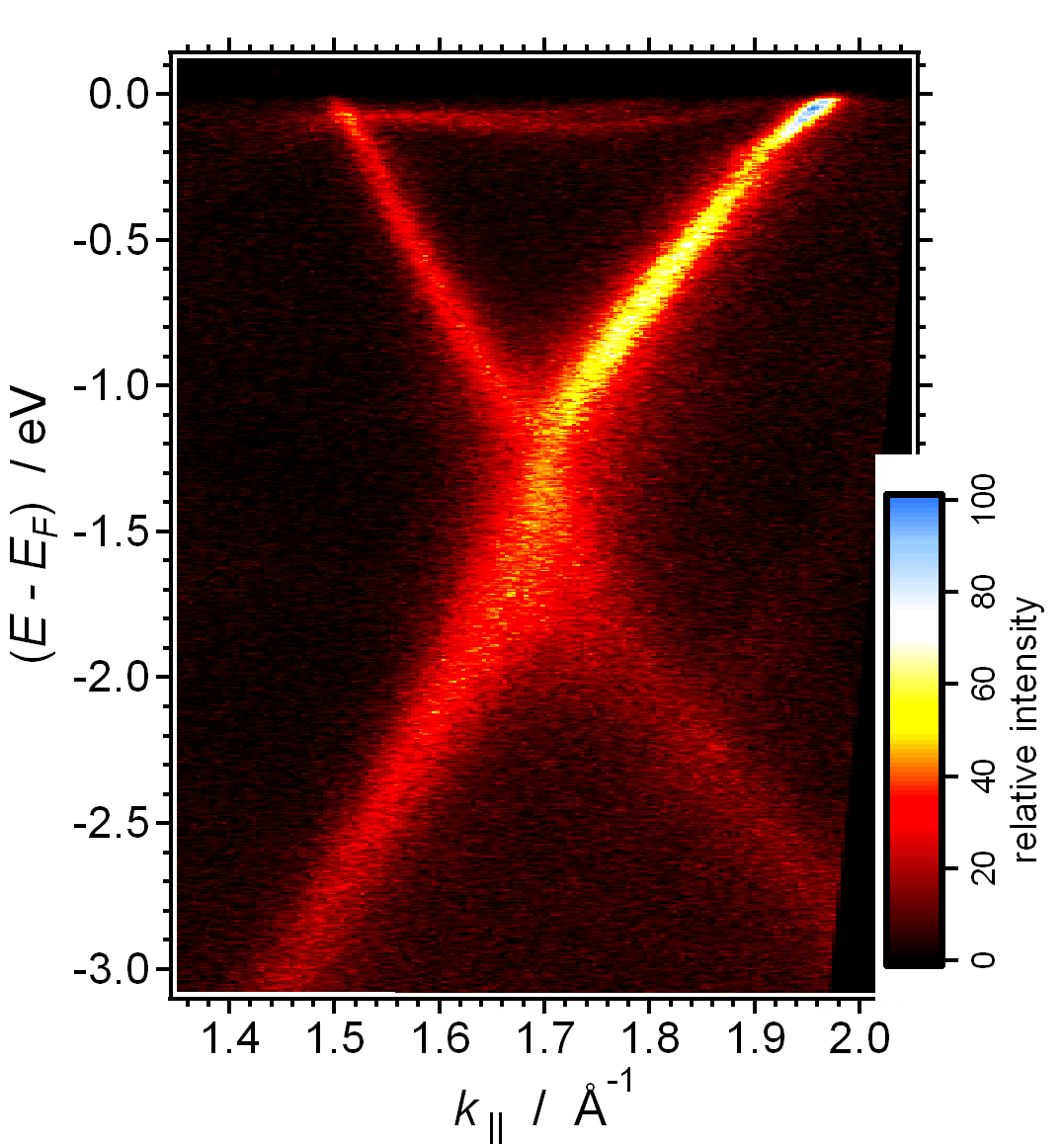}
\par\end{centering}

\caption{\label{fig:ARPES_G-K-M}(Color online) ARPES spectrum (excitation
energy 40.5~eV, $s$ polarization, scan direction $\Gamma$-K-M)
showing the intensity of photoemission from the Dirac cone in graphene/Ir(111)
saturated with potassium.}
\end{figure}
Figure \ref{fig:ARPES_G-K-M} shows photoemission spectrum of potassium-intercalated
graphene on Ir(111) around the K point along the $\Gamma$-K-M direction.
The spectrum shows a discernible asymmetry of the intensity, such
that the part of the spectrum of the $\pi^{*}$ band along K-M ($k>1.7$~Å\textsuperscript{-1})
exhibits much stronger intensity than the part dispersing along K-$\Gamma$
($k<1.7$~Å\textsuperscript{-1}). Notice that iridium surface state
at the Fermi level, also present in the spectrum for graphene/Ir(111),
\citep{Pletikosic2009} is not quenched by the intercalation of potassium.
Potassium doping, however, has rather strong effect on the $\pi$
bands of graphene \textemdash{} shifts the Dirac point to higher binding
energies and renormalizes the band dispersion just below the Fermi
level.

The position of the Dirac point, being defined as a single point in
momentum space where the $\pi$ bands cross each other, is not straightforward
to determine. As can be seen from Fig. \ref{fig:ARPES_G-K-M} there
is no such well-defined crossing point here. This is presumably due
to a band gap, as a study by \citet{Varykhalov2010} has shown that
some metallic dopands do induce a band gap at the Dirac point. We
estimate the width of the band gap to be 0.3~eV with the position
of the Dirac point at 1.35~eV below the Fermi level, which is, within
a tenth of an eV, equal to the values previously obtained for graphene/Ir(111)
\citep{Bianchi2010} and some other systems, such as graphene/SiC
\citep{Gruneis2009} and graphite\citep{Valla2009}.

The shift of the Dirac point to higher binding energies causes an
increase of the Fermi surface, which is for higher doping levels characterized
by trigonal warping; i.~e. an effect when a transformation of the
constant energy maps from circular to trigonal shape takes place.
\citep{Gruneis2009,McChesney2010} The trigonal warping is clearly
associated with different values of $k_{F}$ for the $\pi^{*}$ band
dispersing from K to M or $\Gamma$. We have determined $k_{F}$ along
the directions K-$\Gamma$ and K-M to be 0.18~Å\textsuperscript{-1}
and 0.27~Å\textsuperscript{-1} respectively, and used these values
to estimate the doping level of about 0.05 electrons per unit cell.

\begin{figure}
\begin{centering}
\includegraphics[width=8.6cm]{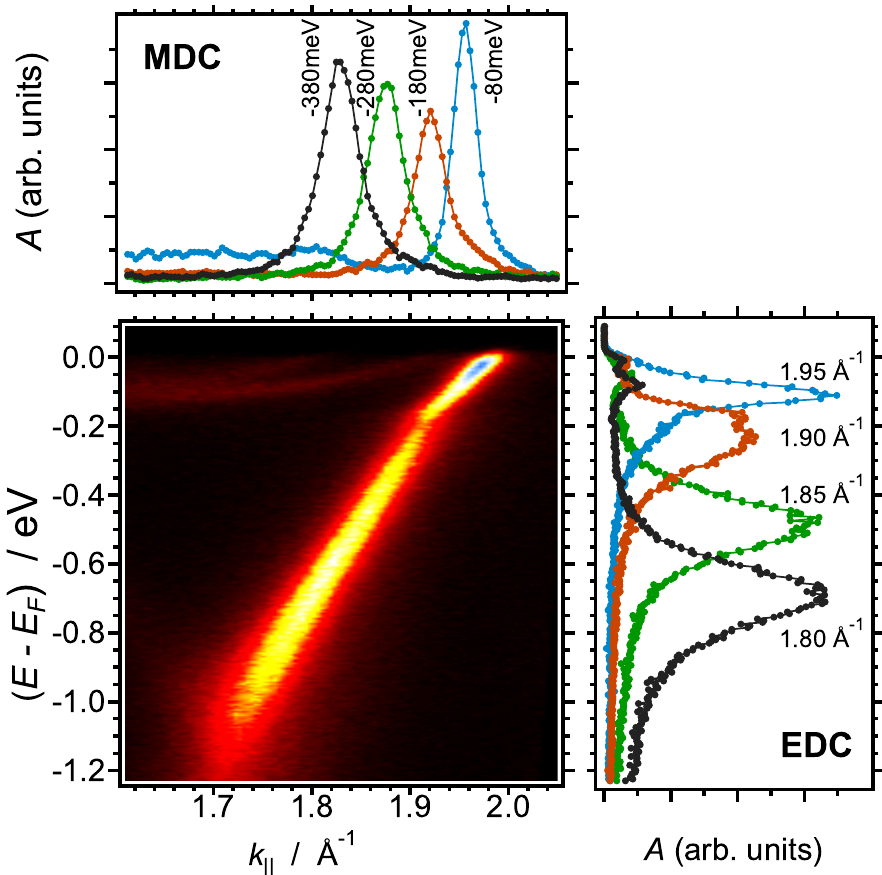}
\par\end{centering}

\caption{\label{fig:ARPES_K-M_MDC_EDC}(Color online) ARPES spectrum of potassium
intercalated graphene/Ir(111) (excitation energy 40.5~eV, \emph{p}
polarization, scan direction $\Gamma$-K-M) along with a few selected
MDC and EDC cuts at given energies and momenta.}
\end{figure}
A prominent feature of the spectrum in Fig. \ref{fig:ARPES_G-K-M}
is the renormalization of the K-M branch of the $\pi^{*}$ band just
below the Fermi level. No such renormalization is as obvious for the
branch K-$\Gamma$. This kink-like change of the band dispersion is
detailed in Fig. \ref{fig:ARPES_K-M_MDC_EDC}. 

As we have pointed out, two parts of the spectrum in Fig. \ref{fig:ARPES_G-K-M}
show a considerable difference of spectral intensity. The change of
the light polarization can additionally alter this intensity ratio.
We used \emph{p} polarization of the incident light in order to extinguish
spectral features along K-$\Gamma$ entirely, leaving only the part
of the $\pi^{*}$ band along the K-M direction visible. The origin
of the diminishing spectral intensity has been explained by \citet{Gierz2011}
The change of the light polarization strongly enhances signal to noise
ratio which shows to be essential in detailed spectrum analysis, as
will be demonstrated in the following. Simple inspection of Fig. \ref{fig:ARPES_K-M_MDC_EDC}
places the phonon induced dispersion kink at around 200~meV below
the Fermi level. Notice how the dispersion kink is now even more pronounced,
accompanied by a strong drop of the spectral intensity at the kink.
In the following we present a detailed analysis of the spectrum along
the K-M direction, and only state the results for the K-$\Gamma$
direction for which the analysis is presented in the supplemental
material, Ref. \onlinecite{SupplemMat}.

\section{Analysis and discussion}

A set of MDC and EDC cuts in Fig. \ref{fig:ARPES_K-M_MDC_EDC} illustrates
the kind of analysis made for each and every slice of the ARPES spectrum
shown. MDCs are generally characterized by a continuous change of
the peak position without any substantial change of the lineshape,
apart for the width of the peaks. However, in the narrow region around
the kink, the EDCs clearly exhibit double peak structure, that can
be fitted with two Lorentzians. The energy splitting between the peaks
is found to be around 60~meV. Farther away, \textpm{}70~meV from
the kink, EDCs can be described by a single Lorentzian. A small peak
visible in EDCs just below the Fermi level is associated with the
S\textsubscript{1} surface state of Ir(111). \citep{Pletikosic2010}

Note that the full width at half maximum of the MDC at an energy just
below the Fermi level (Fig. \ref{fig:ARPES_G-K-M}) is 0.022~Å\textsuperscript{-1},
which is even slightly smaller than the width measured for bare graphene
on Ir(111) \citep{Kralj2011} and comparable to non-intercalated graphene
on SiC \citep{Bostwick2007c}. The measured width of 0.022~Å\textsuperscript{-1}
is substantially smaller than the previously reported value of 0.095~Å\textsuperscript{-1}
for the same system. \citep{Bianchi2010} This questions the conclusion
by \citet{Bianchi2010} that the doping of graphene/Ir(111) by potassium
increases the electron scattering rate due to the loss of translation
symmetry induced by the incommensurability of graphene and Ir(111).
Our data imply that the intercalation of potassium, observed as a
disappearance of graphene\textquoteright{}s moiré superstructure,
does not increase the electron scattering rate compared to bare graphene
on Ir(111). Therefore, we can conclude that potassium intercalated
into graphene/Ir(111) system does not act as an additional scattering
center. As we shall demonstrate later, the measured value of the MDC
width translates into rather big quasiparticle scattering time. The
MDC width increase (to 0.042~Å\textsuperscript{-1}) measured below
the dispersion kink (Fig. \ref{fig:ARPES_G-K-M}) fits into the picture
of phonon-induced renormalization of the $\pi^{*}$ band. 

\begin{figure}
\begin{centering}
\includegraphics[width=8.6cm]{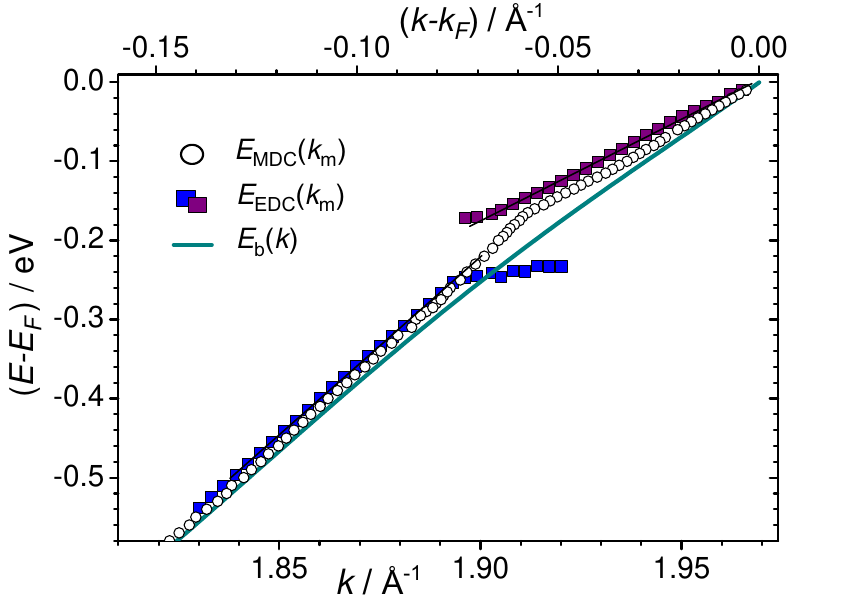}
\par\end{centering}

\caption{\label{fig:MDC_EDC_Eb}(Color online) Peak positions extracted from
MDC cuts (circles) and EDC cuts (squares) of the spectrum shown in
Fig. \ref{fig:ARPES_K-M_MDC_EDC}.  Thick line shows the bare-band
$E_{b}(k)$ obtained from the self-consistent procedure. }
\end{figure}
Figure \ref{fig:MDC_EDC_Eb} summarizes the peak positions obtained
from MDC and EDC sets of spectra from Fig. \ref{fig:ARPES_K-M_MDC_EDC}.
Open circles represent peak positions obtained from the analysis of
MDCs, while squares are obtained from EDCs. The fact that the two
dispersions do not coincide, especially in the region above the kink
is easy to understand when one takes into account that the spectral
intensity of the band sharply changes, and that what is a local maximum
$k_{m}$ in a horizontal cut along $E=E_{m}$ does not have to be
a local maximum in a vertical cut along $k=k_{m}$. The data below
and above the kink are fitted with two linear functions with noticeably
different slopes. The data far away from the Fermi level can be fitted
with a linear curve and the corresponding band velocity equals to
$0.7\cdot10^{6}$~m/s. This represents a reduction of about 30\%
compared to the $\pi$ band of nearly neutral graphene on Ir(111),
and is consistent with the predicted reduction of the Fermi velocity,
as calculated by \citet{Park2009} The set of MDC data between the
Fermi level and the kink is also fitted with a linear function defined
by the band velocity (which is in this case the Fermi velocity as
well) $v_{r}(E_{F})=0.46\cdot10^{6}$~m/s. Note that this value is
only about half the Fermi velocity of the $\pi$ band in bare graphene
on Ir(111) \citep{Kralj2011} and the $\pi^{*}$ band on SiC \citep{Siegel2011}.
This renormalization of the velocity can lead to an overestimate of
the coupling strength to phonons for the simple reason that the change
of the electron velocity near $E_{F}$ is largely due to the curvature
of the $\pi^{*}$ band induced by other interactions and only to small
extent by electron-phonon coupling. \citep{Park2008a}

In order to take into account the nonlinearity of the $\pi^{*}$ band,
the bare-band function $E_{b}(k)$ was self-consistently reconstructed
from the MDC data, by the recursive procedure described above. The
$\pi^{*}$ band itself has been measured far below the phonon-induced
kink, where the contribution to $\Re\Sigma$ can be neglected and
the bare-band approaches the measured. \citep{Kordyuk2005} This gives
confidence to the functions obtained, as the problem of tails is greatly
avoided. The corrections of $E_{m}(k)$ leading to $E_{b}(k)$ being
mostly due to low-energy phonon excitations, the function $E_{b}(k)$
can be considered as one that takes into account all electron interactions
except those with phonons. 

The derivative of the bare function at the Fermi level gives the Fermi
velocity for the bare band $v_{b}(E_{F})=0.56\cdot10^{6}$~m/s. The
obtained values of band velocities, at the Fermi level and far away
from the kink, are in very good agreement with the values of the band
velocities for highly doped graphene, reported by \citet{Siegel2011}

To summarize, it is clear that a large portion of the velocity reduction
as we approach the Fermi level comes from the non-linear nature of
the bare $\pi^{*}$ band. The modification of the Fermi velocity due
to the coupling to phonons is accordingly rather small, around 0.1·10\textsuperscript{6}~m/s,
which is only a 10\% change.

The ratio between the Fermi velocity of the bare band and the electron-phonon-induced
renormalized velocity at the same energy gives, by eq. (\ref{eq:lambdaVbVr}),
the value of the electron-phonon coupling strength. Accordingly, the
value of $\lambda$ along the K-M direction equals to 0.22. The same
analysis on the K-$\Gamma$ direction gives $\lambda=0.19$. \citep{SupplemMat}
Previously, \citet{Bianchi2010} obtained $\lambda=0.28$ for both
directions. 

\begin{figure*}
\begin{centering}
\includegraphics[height=5.3cm]{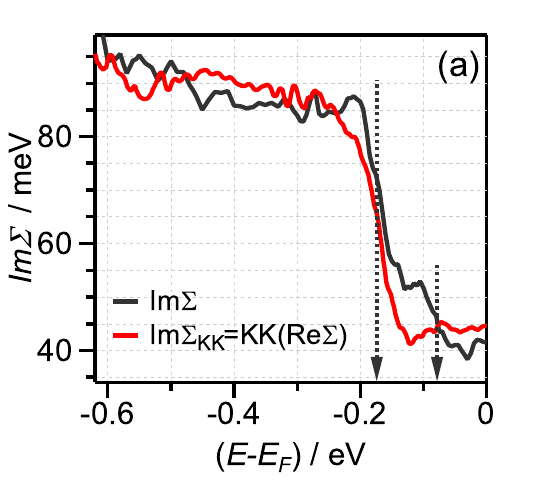}\includegraphics[height=5.3cm]{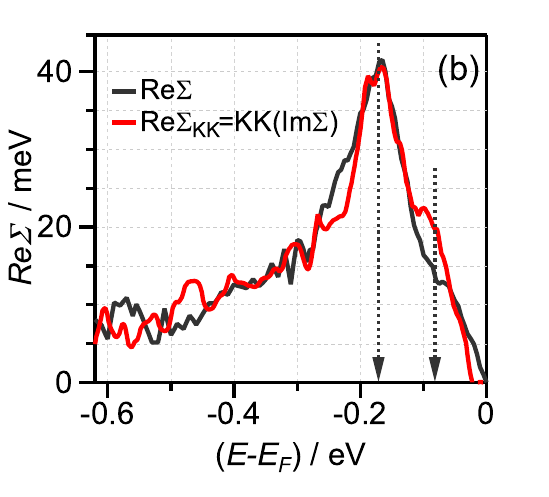}\includegraphics[height=5.3cm]{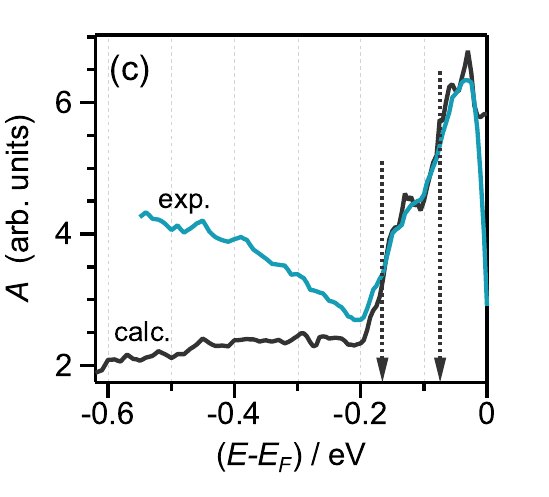}
\par\end{centering}

\caption{\label{fig:ImS_ReS_A}(Color online) A set of self-consistent functions
obtained from MDC cuts of the spectrum shown in Fig. \ref{fig:ARPES_K-M_MDC_EDC}
and the bare-band shown in Fig. \ref{fig:MDC_EDC_Eb}. (a) black line:
$\Im\Sigma$ calculated by eq. (\ref{eq:ImS}); red line: $\Im\Sigma_{KK}$
as a Kramers-Kronig transform of $\Re\Sigma$; (b) black line: $\Re\Sigma$
calculated by eq. (\ref{eq:ReS}); red line: $\Re\Sigma_{KK}$ as
a Kramers-Kronig transform of $\Im\Sigma$. (c) spectral intensity
$A(E)$ from the experiment (blue line) and calculated from the self-consistent
data by eq. (\ref{eq:A}) (black line). Arrows indicate the energies
of phonons that induce band renormalization.}
\end{figure*}
We use MDCs extracted from the spectrum shown in Fig. \ref{fig:ARPES_K-M_MDC_EDC}
to plot (a) imaginary part of the self-energy, $\Im\Sigma(E)$ (b)
real part of the self-energy, $\Re\Sigma(E)$ and (c) spectral intensity
of the MDC peaks, $A(E)$. Figure \ref{fig:ImS_ReS_A} shows the results.
The functions plotted in black were calculated from formulæ (\ref{eq:A})--(\ref{eq:ImS})
using the bare-band from our self-consistent iterative procedure.
The functions shown in red illustrate the degree of self-consistency
achieved; the one for $\Im\Sigma_{\mathrm{KK}}$ in Fig. \ref{fig:ImS_ReS_A}(a)
was obtained by Kramers-Kronig transformation of the data for $\Re\Sigma$
from Fig. \ref{fig:ImS_ReS_A}(b), and equally, the one for $\Re\Sigma_{\mathrm{KK}}$
from the data for $\Im\Sigma$. The overall agreement is quite good.
We find the data for FWHM and consequently for $\Im\Sigma$ more reliable,
the noise being pretty low and the relative error being smaller in
determining the width of the MDCs than the peak positions, from which
$\Re\Sigma$ is calculated. The effects of experimental smearing also
have a greater impact on the fine details of MDC peak positions, adding
only a constant background to FWHMs and $\Im\Sigma$. We suspect that
the low-energy shoulder in $\Re\Sigma$ (and the associated step in
$\Im\Sigma_{KK}$) are for these reasons less pronounced than the
corresponding ones in $\Re\Sigma_{KK}$ (and $\Im\Sigma$). The difference
in experimental and calculated spectral intensities in Fig. \ref{fig:ImS_ReS_A}(c)
at the high-energy side probably comes from the fact that the calculated
renormalized band $E_{b}(k_{m})+\Re\Sigma(E_{m})$ does not come close
enough to the measured band $E_{m}(k_{m})$, as the increase of the
spectral function would come from one approaching the other (see Eq.
\ref{eq:A}). We suspect this could be a still-observable consequence
of the missing high-energy tails. \citep{Kordyuk2005}

Generally, a step-like increase of $\Im\Sigma(E)$ and a maximum of
$\Re\Sigma(E)$ at the same energy is associated with electron coupling
to some excitation, in this case phonons. As pointed out by \citet{McChesney2008}
the coupling to a phonon should have a profound impact on the energy
dependence of the spectral intensity $A(E)$ as well. 

Instead of one, all displayed spectral parameters show consistently
two features that can be accordingly associated with the coupling
to two phonons, one at 170~meV and the second one at 75~meV. Arrows
in Figs. \ref{fig:ImS_ReS_A} (a)--(c) indicate the energies of these
two phonons.

The $\Im\Sigma(E)$ (Fig. \ref{fig:ImS_ReS_A}a) shows a distinct
step-like increase between 0.05 and 0.1~eV below the Fermi level
with the initial value of 39~meV which reaches a local maximum of
52~meV at 0.1~eV. This increase of $\Im\Sigma(E)$ is associated
with the phonon of energy 75~meV. Further increase of $\Im\Sigma(E)$
up to 85~meV induced by the coupling to a phonon of 170~meV takes
place between 0.13 and 0.21~eV. Below that, $\Im\Sigma(E)$ stays
constant up to 0.4~eV, but then starts to increase with energy, due
to other contributions to the lifetime (electron-electron, electron-plasmon).
\citep{Bostwick2007c}

Electron-phonon coupling constant can be extracted from the change
of $\Im\Sigma(E)$ by eq. (\ref{eq:lambdaIm}). \citep{Fink2006,Gruneis2009}
The total increase of $\Im\Sigma(E)$ equal to 46~meV, translates
to $\lambda=0.17\pm0.01$ along the K-M direction. A similar analysis
of $\Im\Sigma(E)$ along the K-$\Gamma$ direction gives $\lambda=0.18\pm0.02$.
\citep{SupplemMat}

$\Re\Sigma(E)$, consistently with $\Im\Sigma(E)$ and $A(E)$, shows
a peak at 170~meV and a distinct shoulder at a lower binding-energy
side, which clearly implies an existence of two excitations that contribute
to the renormalization of the bare band dispersion. 

Using the imaginary part of the self-energy in the proximity of the
Fermi level, we determine the photohole scattering time, $\tau=\frac{\hbar}{2\,\Im\Sigma}$,
\citep{Calandra2007a} to be bigger than 10 fs. This is the same value
as obtained for high quality graphene on SiC. \citep{Sprinkle2009}
Note that this value is not to be compared to the scattering time
in mobility measurements ($\sim350$~fs), as those include some more
scattering mechanisms. \citep{Calandra2007a}

As pointed out, the energy dependence of the spectral intensity $A(E)$
also indicates the existence of two distinct phonon excitations that
couple to electrons. \citet{McChesney2008} have demonstrated a high
sensitivity of $A(E)$ to many-body interactions, drawing attention
to the fact $A(E)$ can access even a faint contributions to the self-energy
with a sensitivity even better than $\Im\Sigma(E)$ can provide. In
agreement with this model we observe two shoulders in $A(E)$ where
an onset of each shoulder corresponds to a phonon excitation (see
Fig. 2 in Ref. \onlinecite{McChesney2008}). Summarizing the features
that have been observed in three parameters: $\Im\Sigma(E)$, $\Re\Sigma(E)$
and $A(E)$, we can conclude, with a high degree of consistency, that
the phonon energies that induce the renormalization of the dispersion
of the $\pi^{*}$ band of doped graphene along the K-M direction have
energies equal to 170~meV and 75~meV. These two phonons can be associated
with an optical phonon (transverse or longitudinal) around K and an
acoustic phonon, respectively. According to \citet{Gonzalez2009},
in non-interacting graphene these phonons should be in-plane oscillations
as the symmetry does not allow coupling of the out of plane phonons
to electrons in graphene. However the presence of the substrate might
break the 2D symmetry of graphene and hence allow the coupling of
both in-plane and out-of-plane oscillations to electrons in graphene. 

Some theoretical calculations also support the notion of a two-phonon-spectrum
that induces renormalization of graphene bands close to the Fermi
level. According to \citet{Calandra2007a}, the $\pi^{*}$ band of
doped graphene should show two kinks in the dispersion (accordingly
accompanied by two steps in the MDC linewidth), one at 195~meV being
attributed to A\textsubscript{1} mode and another at 160~meV which
corresponds to a twofold degenerate E\textsubscript{2g} mode. Previous
measurements of \citet{Bianchi2010} with graphene on Ir(111) showed
signatures in the band dispersion and widths of spectral curves of
apparently only one phonon. However, their self-consistent analysis
showed that the experimental data can be modelled by contributions
of five Einstein oscillators with energies that are evenly distributed
over the range from 21~meV to 190~meV. 

Interestingly, the pattern of coupling similar to the one we observed
in graphene/Ir(111) is also reported for CaC\textsubscript{6} where
the contributions to the self-energy come from two phonons, one at
75~meV and the other at around 160~meV. \citep{Valla2009} This
agreement supports the notion that intercalated atoms (K or Ca) do
not participate in the coupling of the $\pi^{*}$ band to phonon modes
apart from the doping effect.

The energy splitting between the EDC peaks has already been observed
in materials with strongly coupled electrons and phonons. \citep{Hengsberger1999,LaShell2000}
 Although the coupling in graphene is not as large, its two-dimensionality
is, according to \citet{Badalyan2012}, a possible reason for an enhancement
of the effective coupling in the vicinity of the phonon energy. An
electron-phonon complex quasiparticle forms, giving rise to a strong
modification of the Dirac spectrum, seen as the branches of the $\pi^{*}$
band in the neighborhood of the phonon emission threshold. 

The value of the electron-phonon coupling constant obtained in this
work ($\sim0.2)$ is somewhat smaller than the one obtained by \citet{Bianchi2010}
for the same system ($\sim0.3)$. Nevertheless, the value derived
from our spectra does not support the notion of \citet{Siegel2011a}
that the coupling in \emph{n}-doped graphene on a metallic surface
should be still a few times smaller ($\sim0.04$). Given sharp enough
spectra, low in background intensity and noise, a pronounced kink
in the $\pi^{*}$ band dispersion, and especially a steplike increase
of the spectral width at the phonon energy, must alone be a clear
sign of a non-negligible strength of interaction, whatever the choice
for the bare band is.

\section{Conclusions}

In conclusion, we have analyzed various parameters from high resolution
ARPES measurements --- peak positions and widths of MDCs, imaginary
and real part of the self-energy $\Sigma(E)$, spectral intensity
$A(E)$ --- all consistent with the notion that two phonons contribute
to the renormalization of the band dispersion in graphene on Ir(111):
a high energy phonon at 170~meV and a low energy phonon at 75~meV.
The coupling constant associated with these two phonons is around
0.2, which is similar to a previous finding for the same system. We
have also found that the intercalation of potassium up to saturation
does not increase the scattering rate of the photohole in the $\pi^{*}$
band. Due to the perfectness of the K/graphene/Ir(111) structure 
the quasiparticle (photohole) scattering time can be posed exceptionally
high, above 10~fs.
\begin{acknowledgments}
The authors thank Alpha N'Diaye and Carsten Busse for the help in
preparation of graphene samples, as well as Ivana Vobornik and Jun
Fujii for the technical assistance at the Advanced Photoemission beamline
at ELETTRA. We gratefully acknowledge financial support by MZOS via
the project No. 035-0352828-2840.
\end{acknowledgments}
\bibliographystyle{apsrev}
\bibliography{graphene-renorm,notes}

\cleardoublepage{}

\begin{spacing}{1.3}
\onecolumngrid
\raggedbottom
\end{spacing}

\appendix
\begin{spacing}{1.3500000000000001}
{\large 
\title{Finding the bare band: Electron coupling to two phonon modes in potassium-doped
graphene on Ir(111) }

\author{I. Pletikosi\'{c}, M. Kralj, M. Milun, P. Pervan}

\affiliation{Institut za fiziku, Bijeni\v{c}ka 46, HR-10000 Zagreb, Croatia }

\maketitle

\section*{Supplemental Material}

\subsection*{Formulæ}

Under the assumption that the self-energy does not depend on momentum,
equations (2) and (3) from our article are exact whatever the bare-band
function $E_{b}(k)$ is. However, in order to use the linearized version
of eq. (3) --- $\Im\Sigma(E_{m})=\hbar v_{b}(E_{m})\cdot w_{m}$,
where $v_{b}(E_{m})=\frac{1}{\hbar}\frac{d}{dk}E_{b}(k)|_{k=k_{m}(E_{m})}$
--- the measured band should be sufficiently sharp ($w_{m}$ sufficiently
small), and the bare band piecewise linear (giving a symmetric peak,
$w_{L\, m}=w_{R\, m}=w_{m}$). Even in our case of very narrow MDC
peaks (0.010~Å\textsuperscript{-1}<$w_{m}$<0.022~Å\textsuperscript{-1},
or 0.007~Å\textsuperscript{-1}<$w_{m}$<0.018~Å\textsuperscript{-1}
when the experimental resolution is deconvolved) this is only partially
fulfilled. The figure below illustrates the relations for $\Im\Sigma$,
$\Re\Sigma$ and $E_{b}$ in the case of real experimental data on
$E_{m}(k_{m})$ and FWHM ($2\, w_{m}$). Also shown is an MDC curve
for $E_{m}=-0.13$~eV with FWHM of 0.026~Å\textsuperscript{-1}.

In the following, we shall not use the linearized version of eq. (3),
in spite of rather narrow peaks, and only make use of $w_{L\, m}=w_{R\, m}=w_{m}$,
as our Lorentzians are quite symmetric.

\bigskip{}

\begin{center}
\includegraphics[width=12cm]{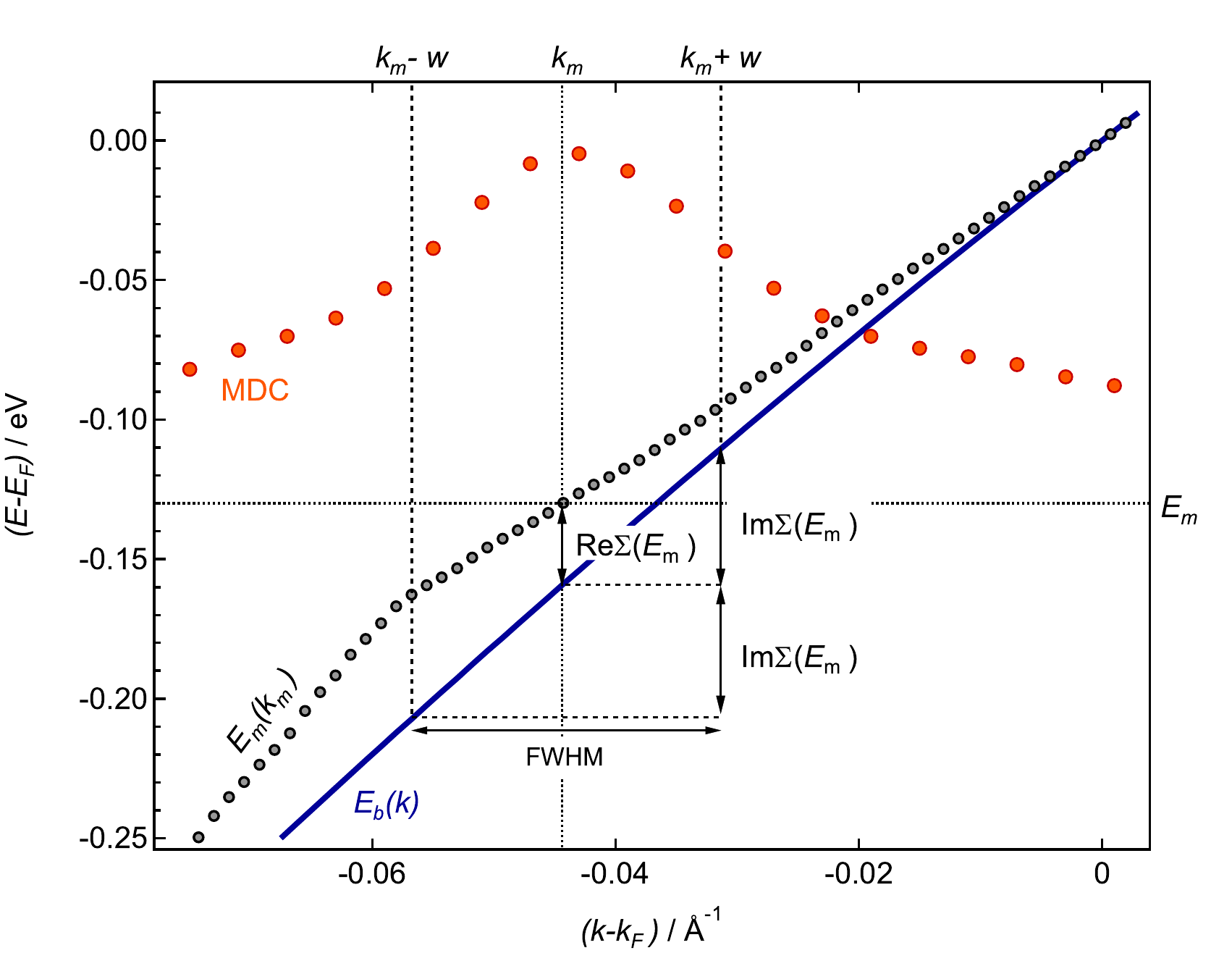}
\par\end{center}

\newpage{}

\subsection*{Method}

\begin{minipage}[b][150mm][t]{75mm}%
Instead of finding $E_{b}(k)$ by cleverly trying the parameters that
define the function in a method commonly used before (Method I on
the right), using Levenberg\textendash{}Marquardt algorithm, for example,
the idea is to employ a loop algorithm that happens to be self-correcting
and relatively easy to implement (Method II on the right).

Here, $\Re\Sigma(E_{m})$ obtained from the Kramers-Kronig transform
of $\Im\Sigma$ is used to calculate $E_{b}$ from the measured renormalized
dispersion $E_{m}(k_{m})$ on a discrete set of points. Then, by least-squares
fitting, a functional form $E_{b}(k)$ is acquired, and used to calculate
$\Im\Sigma$ in the next iteration.%
\end{minipage} \hspace{8mm}%
\begin{minipage}[b][150mm][t]{85mm}%
\begin{center}
\textcolor{red}{METHOD I}
\par\end{center}

\smallskip{}

\includegraphics[width=80mm]{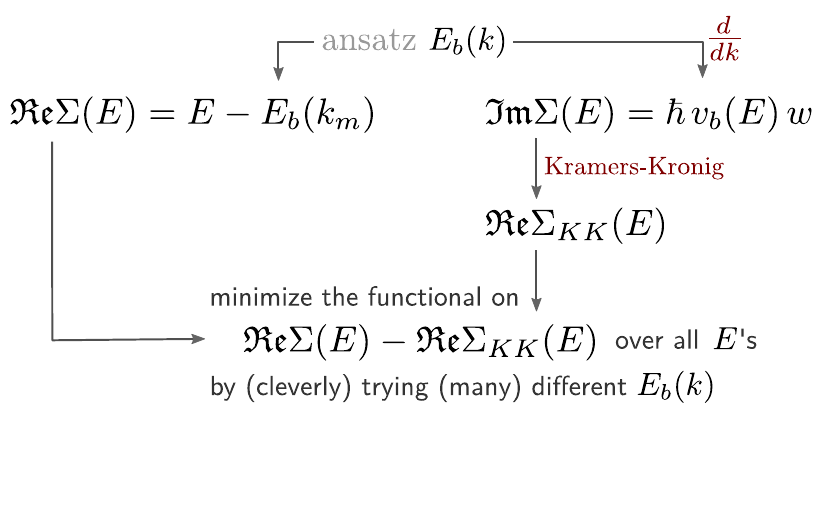}

\bigskip{}

\begin{center}
\textcolor{red}{METHOD II }
\par\end{center}

\smallskip{}

\includegraphics[width=80mm]{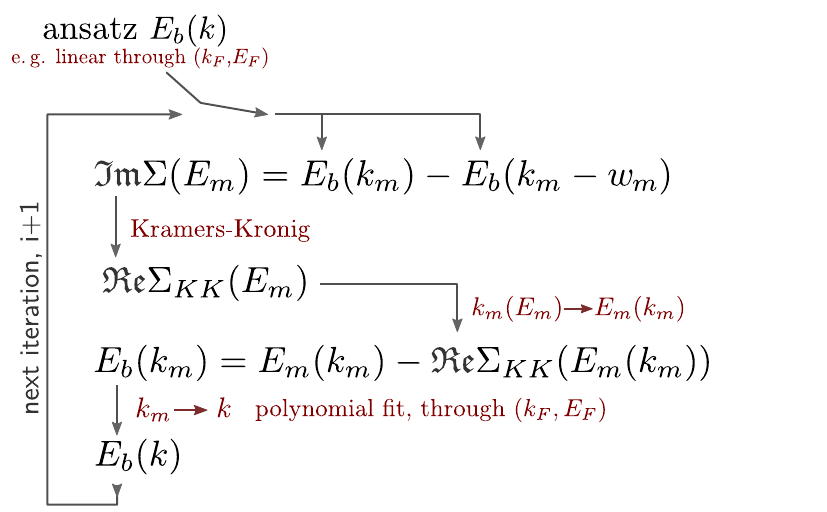}%
\end{minipage} 

\newpage{}

\subsection*{Convergence}

The figure below shows the convergence of the bare-band function.
The iteration starts by postulating a linear band (i\textsubscript{o}).
Already the second iteration (i\textsubscript{2}) gives the band
indistinguishable from those obtained in the following iterations.
Also shown is how the coefficients of a third-order polynomial for
$E_{b}$

\[
E_{b}(k)=0+c_{1}k+c_{2}k^{2}+c_{3}k^{3}
\]
converge oscillatory through the course of iterations.

Notice that the renormalized and the bare-band merge quite close to
the kink, giving confidence that the influence of the missing tail
is negligible.

\bigskip{}

\begin{center}
\includegraphics[width=15cm]{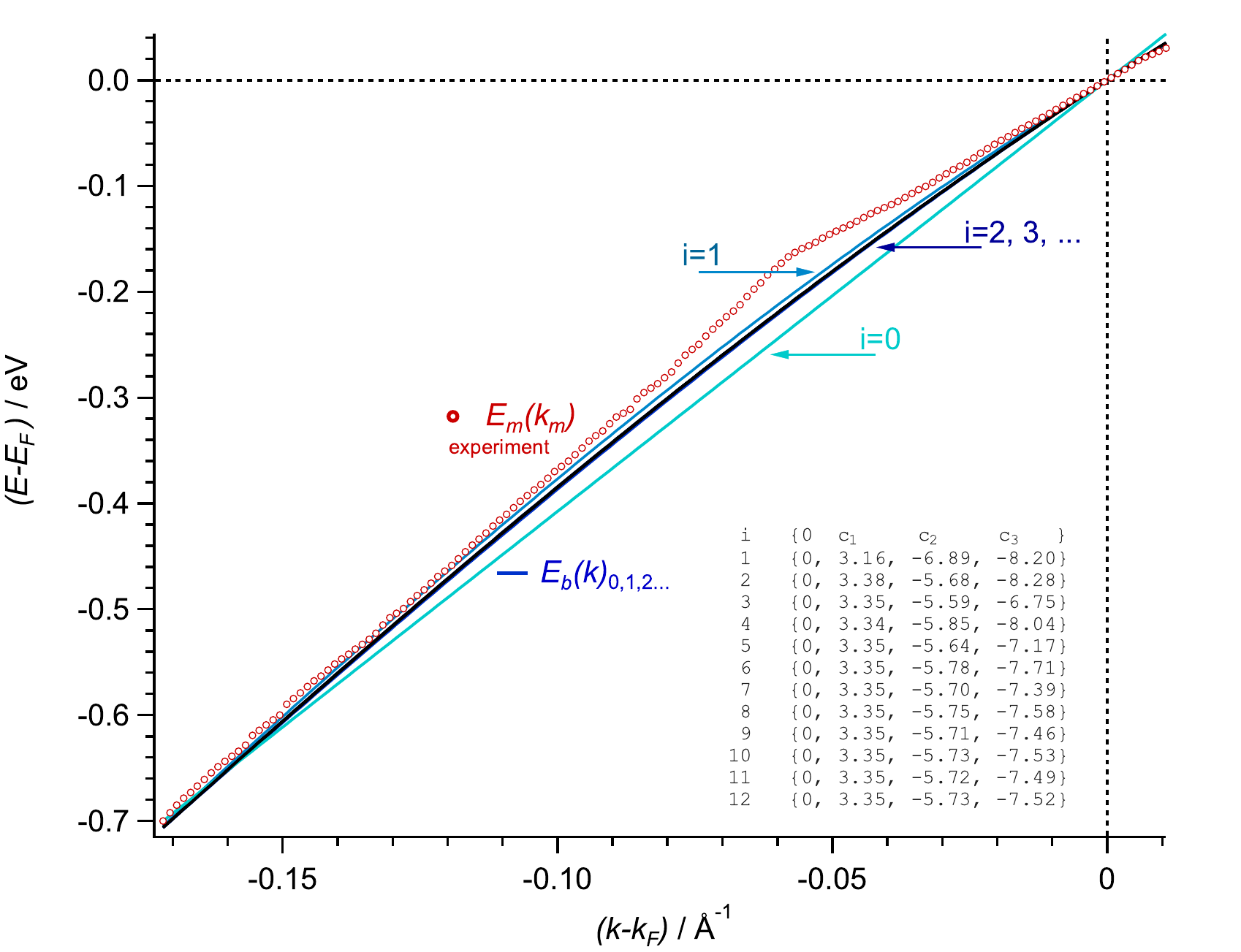}
\par\end{center}

\newpage{}

\subsection*{Code}

Two procedures to be used in Wavemetrics IgorPro data analysis software
are presented. The first procedure, \texttt{\footnotesize scKK\_setup},
is used for initialization, with properly scaled one-dimensional waves
containing data on the positions (\texttt{\footnotesize km}) and widths
(\texttt{\footnotesize fwhm}) of MDC Lorentzian curves; examples are
given in the next section. The second procedure, \texttt{\footnotesize scKK\_iteration},
is called for several times, until the convergence in the output waves
(\texttt{\footnotesize scEb}, \texttt{\footnotesize scEb\_fit}, \texttt{\footnotesize scImS},
\texttt{\footnotesize scReS}\ldots{}) is achieved.

We find it convenient to put ($k_{F}$,$E_{F}$)=(0,0), as that point
can be read-off the spectrum with high enough accuracy, and make the
bare-band function pass through it.

Instead of a polynomial, any fitting function (e.g. a tight-binding
formula) should be easy to implement. As we are usually not interested
in $E_{b}$, but in the results obtained for $\Im\Sigma$ and $\Re\Sigma$,
polynomial fitting is sufficient.

\vspace{1cm}

\shadowbox{\begin{minipage}[t]{0.9\textwidth}%
\begin{lstlisting}[basicstyle={\tiny\ttfamily},commentstyle={\color{darkred}},morecomment={[l]{//}},showstringspaces=false,tabsize=4]

//
// Author: Ivo Pletikosic, ivo.pletikosic@ifs.hr, November 2011
//
// sc = self-consistent;    KK=Kramers-Kronig



// first, create waves needed for the iteration procedure
// wave km(Em) contains the experimental dispersion, i.e. maxima of the spectral function
// wave fwhm(Em) contains widths of Lorentzian peaks at a given energy
// x axes should be properly scaled!
// energy is measured from the Fermi level, (E-EF)
Function scKK_setup(km,fwhm)
    Wave km,fwhm

        // (km-kF) as a function of Em
    Duplicate/O km,scKKm
        // putting the origin of the k axis to kF (a point where Em=0)
    Variable kF=km(0);  scKKm=km-kF
    Duplicate/O fwhm,hwhm;     hwhm*=0.5
        // sample ReS at the same energies as KKm
    Duplicate/O scKKm,scReS
        // sample ImS at the same energies as hwhm
    Duplicate/O hwhm,scImS
        // discrete bare band Eb; Em(km); and an auxiliary wave
    Make/O/N=(numpnts(scKKm)) scEb,scEm,Em_scaling
        // scaling from the first/last k-point of the experimental dispersion
    SetScale/I x,scKKm[0],scKKm[numpnts(scKKm)], "A\S-1\M",scEb, scEm
    SetScale  y,0,0,"eV",scEb,scEm
        // this will be used in the interpolation to get Em(km) from km(Em)
    SetScale/I x,leftx(scKKm),leftx(scKKm)+deltax(scKKm)*(numpnts(scKKm)-1),Em_scaling
    Em_scaling=x
        // Em as a function of (km-kF)
    scEm=interp(x,scKKm,Em_scaling)

    // zeroth approximation to Eb:

        // a line through (kF,EF)=(0,0) and the lowest point of the experim. dispersion
    Make/O cw={0,leftx(scKKm)/scKKm[0]}
    scEb=poly(cw,x)
        // this will become symmetrized and resampled ImS
    Make/O/N=(4*numpnts(scImS)) symmImS
    SetScale/I x,leftx(scImS),-leftx(scImS),"eV",symmImS
        // this will become symmetrized and resampled ReS
    Make/O/N=(4*numpnts(scReS)) symmReS
    SetScale/I x,leftx(scReS),-leftx(scReS),"eV",symmReS
End
\end{lstlisting}
\end{minipage}}

\newpage{}

\shadowbox{\begin{minipage}[t]{0.9\textwidth}%
\begin{lstlisting}[basicstyle={\tiny\ttfamily},commentstyle={\color{darkred}},morecomment={[l]{//}},showstringspaces=false,tabsize=2]
// iteration function
// call it several times until there's no substantial change in the output waves
// polyN is the degree of the band fitting polynomial [2, 3, ..., 7]

Function scKK_iteration(polyN)
    Variable polyN
    Wave scKKm, scEb, scReS, scImS
    Wave Em, hwhm, cw, symmReS, symmImS // some auxiliary waves

    PauseUpdate;

        // start from ImS
    scImS=poly(cw,scKKm(x)) - poly(cw,scKKm(x)-hwhm(x))
        // or
    //scImS=poly(cw,scKKm(x)+hwhm(x)) - poly(cw,scKKm(x))
        // or
    //scImS=0.5*(poly(cw,scKKm(x)+hwhm(x)) - poly(cw,scKKm(x)-hwhm(x)))

    // K-K transform ImS to HReS

        // symmetrize
    symmImS=(x<=0) ? scImS(x) : scImS(-x)
        // ReS from the K-K transform of ImS
    HilbertTransform/DEST=scHReS symmImS
    scHReS*=-1
    SetScale/P x,leftx(symmImS),deltax(symmImS),"eV",scHReS

    // ansatz for Eb in the next iteration

        // bare band as a function of km (therefore, of all the discrete k's present in the spectrum)
    scEb=Em(x)-scHReS(Em(x))

        // fit Eb to a polynomial (could be any other function), as we'll need some extrapolation

        // initial values of the fitting polynomial coefficients: all 0
    Make/O/N=(polyN+1) cw=0
        // keep c0=0, i.e. make the polynomial zero at (kF,EF)=(0,0)
    String fmtStr;        sprintf fmtStr,"1%0*.0f" polyN, 0;
        // functional (i.e. smooth) form of the bare band
    Duplicate/O scEb,scEb_fit
        // the output of this cmd will be used for the next iteration
    CurveFit/Q/H=fmtstr/NTHR=1 poly (polyN+1), kwCWave=cw, scEb /D=scEb_fit
        // print fitting coefficients
    print cw

    // and just to check things out
        // ReS from the very definition
        // poly(...) represents a smooth function for Eb from the previous iteration
    scReS=x - poly(cw,scKKm(x))
        // antisymmetrize
    symmReS=(x<=0) ? scReS(x) : (-scReS(-x))
        // ImS from the K-K transform of ReS
    HilbertTransform/DEST=scHImS symmReS
    SetScale/P x,leftx(symmReS),deltax(symmReS),"eV",scHImS
        // offset, as HilbertTransform seems to make mean=0
    scHImS+=mean(scImS)
End
\end{lstlisting}
\end{minipage}}

\newpage{}

\subsection*{Data}

Examples of input waves for the self-consistent iterative procedure,
as extracted from MDCs of an ARPES spectrum analyzed in our paper.

\vspace{10mm}

IgorPro file \texttt{\textcolor{red}{km.itx}}

(here downsampled to $\Delta=0.012$~eV from the original $\Delta=0.005$~eV)

\medskip{}

\shadowbox{\begin{minipage}[c][75mm]{0.44\textwidth}%
\begin{lstlisting}[basicstyle={\tiny\ttfamily},showstringspaces=false,tabsize=6]
IGOR
WAVES    km
BEGIN
    1.797    1.799    1.801    1.804    1.806
    1.809    1.812    1.814    1.817    1.819
    1.822    1.824    1.827    1.830    1.834
    1.836    1.838    1.841    1.844    1.846
    1.849    1.852    1.854    1.857    1.859
    1.862    1.864    1.867    1.870    1.872
    1.875    1.878    1.880    1.883    1.886
    1.889    1.891    1.893    1.895    1.898
    1.900    1.903    1.905    1.907    1.910
    1.913    1.918    1.922    1.927    1.932
    1.936    1.940    1.944    1.947    1.951
    1.955    1.959    1.963    1.968    1.971
    1.975
END
X SetScale/P x -0.7,0.012,"eV", km
X SetScale y 0,0,"1/A", km
\end{lstlisting}
\end{minipage}}\quad{}%
\begin{minipage}[b][70mm][c]{0.45\textwidth}%
\includegraphics[width=1\textwidth]{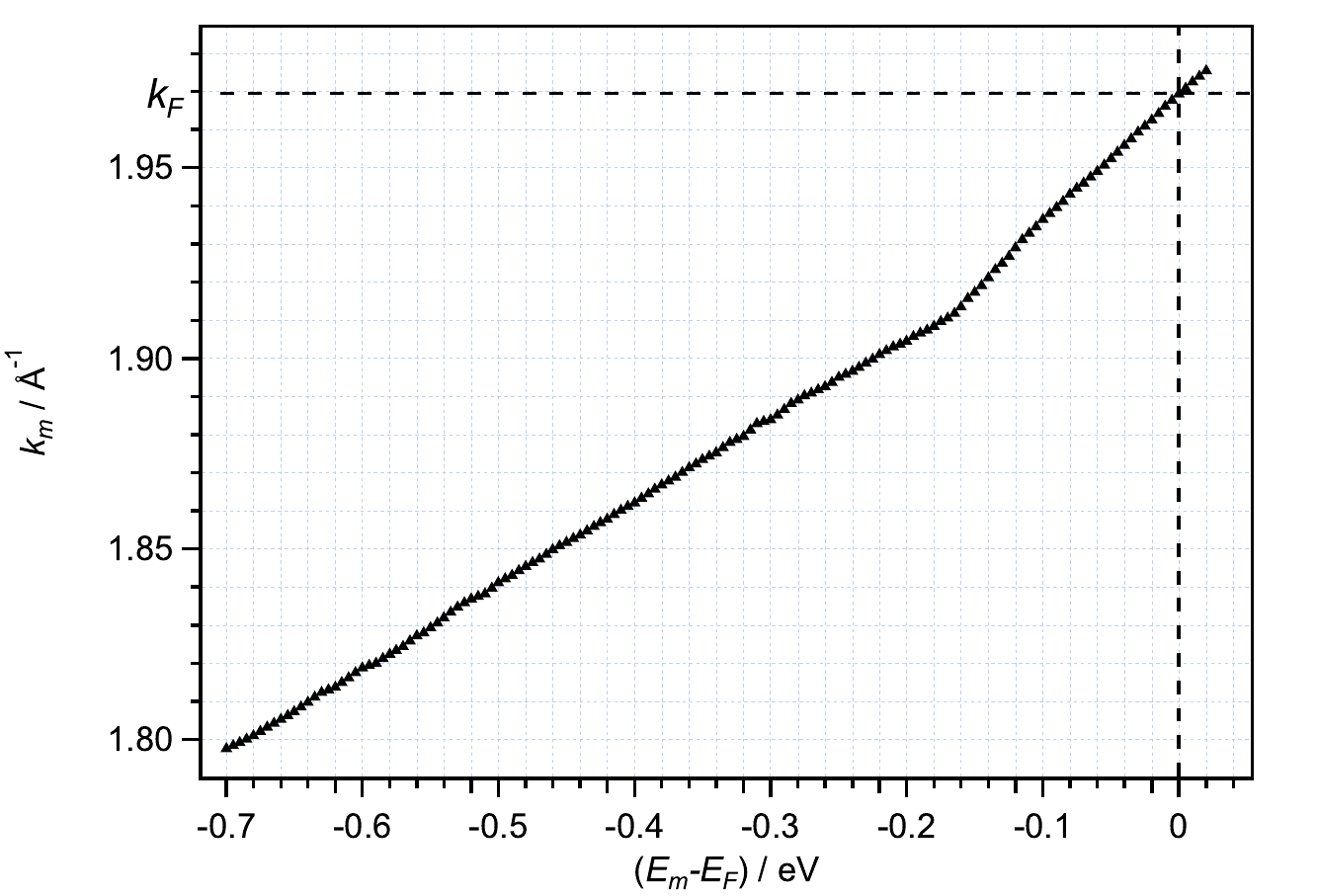}%
\end{minipage}

\vspace{6mm}

IgorPro file \texttt{\textcolor{red}{fwhm.itx}}

(here downsampled to $\Delta=0.012$~eV from the original $\Delta=0.005$~eV)

\medskip{}

\shadowbox{\begin{minipage}[c][75mm]{0.44\textwidth}%
\begin{lstlisting}[basicstyle={\tiny\ttfamily},showstringspaces=false]
IGOR
WAVES    fwhm
BEGIN
    0.0464    0.0458    0.0453    0.0434    0.0430
    0.0445    0.0435    0.0442    0.0423    0.0414
    0.0420    0.0407    0.0418    0.0419    0.0415
    0.0405    0.0414    0.0420    0.0416    0.0408
    0.0398    0.0390    0.0400    0.0405    0.0408
    0.0397    0.0398    0.0399    0.0403    0.0405
    0.0408    0.0408    0.0413    0.0406    0.0399
    0.0419    0.0421    0.0411    0.0415    0.0418
    0.0433    0.0436    0.0428    0.0382    0.0361
    0.0315    0.0290    0.0282    0.0273    0.0279
    0.0278    0.0262    0.0243    0.0233    0.0230
    0.0228    0.0224    0.0245    0.0247    0.0246
    0.0222
END
X SetScale/P x -0.7,0.012,"eV", fwhm
X SetScale y 0,0,"1/A", fwhm
\end{lstlisting}
\end{minipage}}\quad{}%
\begin{minipage}[b][70mm][c]{0.45\textwidth}%
\includegraphics[width=1\textwidth]{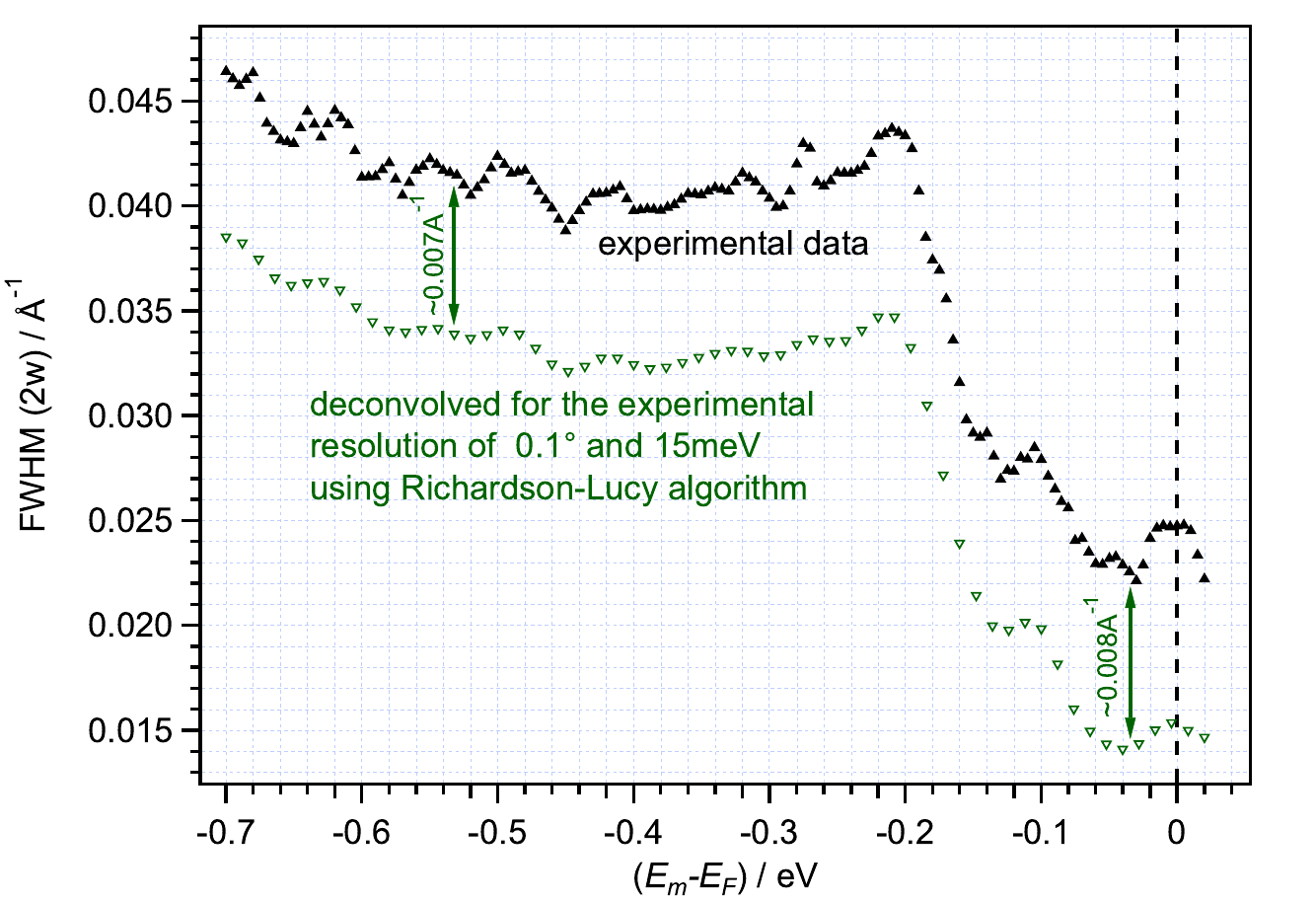}%
\end{minipage}

\newpage{}

\subsection*{A few questions to answer }
\begin{itemize}
\item \emph{Could the $\mathrm{K}$-$\Gamma$ direction be treated with
the same rigor as the }K-M\emph{ direction?}
\end{itemize}
In the experimental geometry we had, the K-$\Gamma$ direction ($k_{||}$<1.7~Å\textsuperscript{-1})
is not visible in high-intensity spectra recorded with the \emph{p}
polarization of light. This is illustrated in the following two raw
images taken with the same excitation energy (40.5~eV) and in the
same energy and emission-angle window. One can easily notice a high
noise level in the \emph{s} polarization spectrum where a part of
the K-$\Gamma$ direction is seen. The photoemission intensity is
about 30 times lower in the \emph{s} polarization spectrum.

\vspace{1cm}

\includegraphics[scale=0.7]{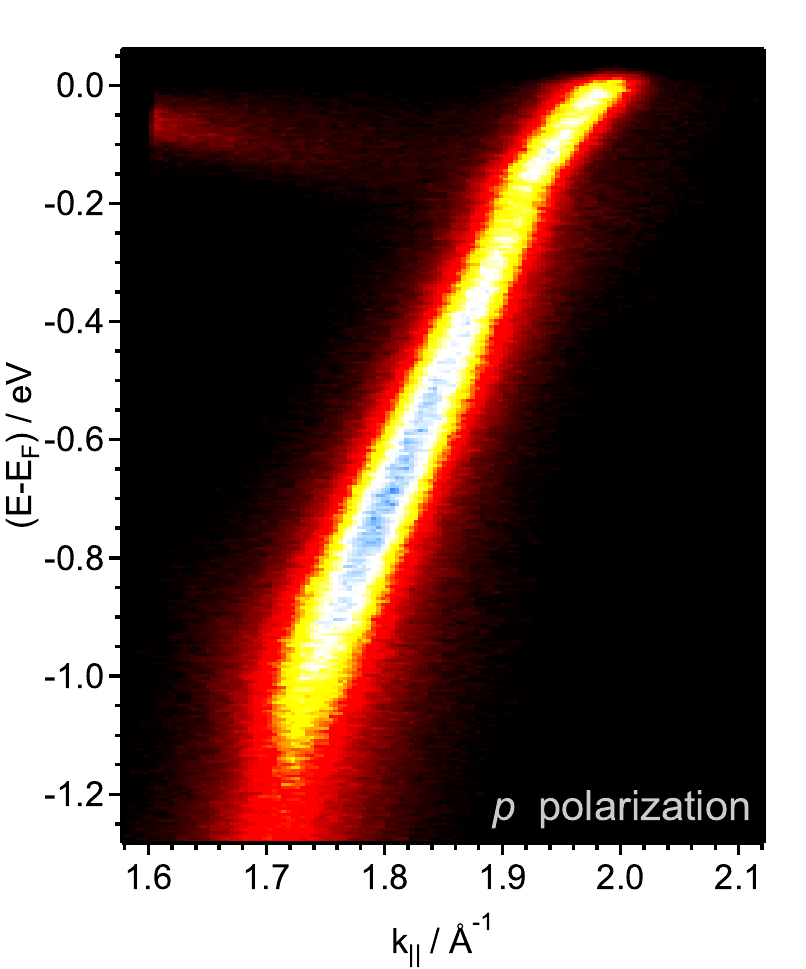}\qquad{}\includegraphics[scale=0.7]{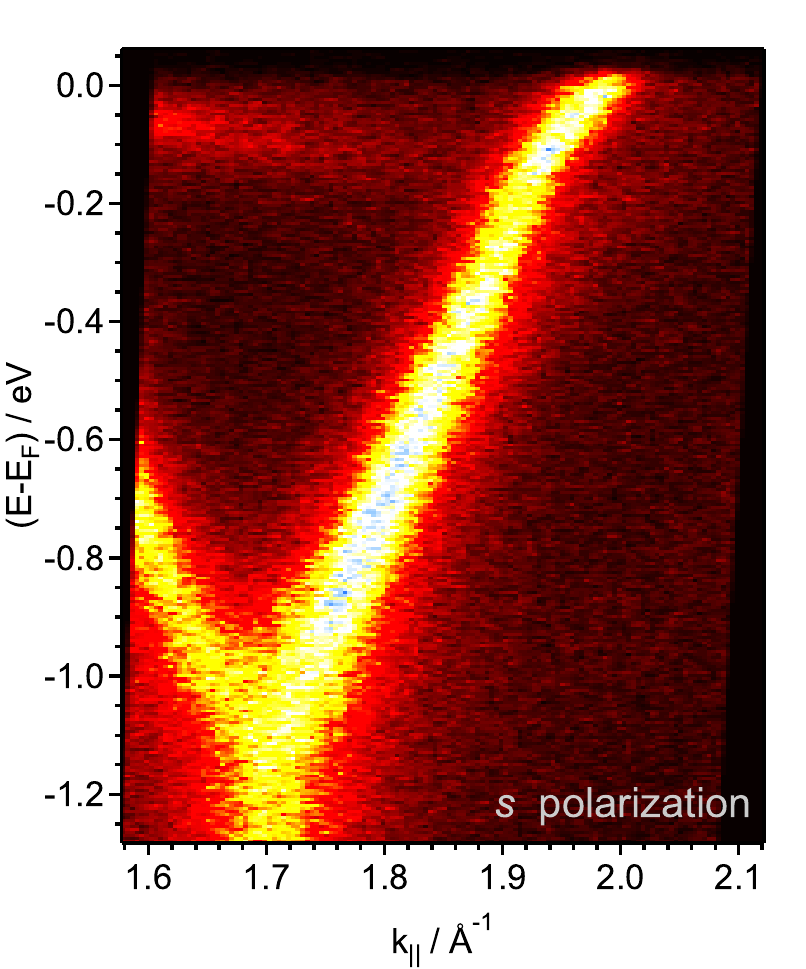}

\newpage{}
\begin{itemize}
\item \emph{So, what are the results for the} \emph{$\mathrm{K}$-$\Gamma$
direction?}
\end{itemize}
We shall analyze the spectrum recorded in the \emph{s} polarization
of the excitation light, shown below. The whole K-$\Gamma$ branch
of the doped $\pi^{*}$ band is visible, continuing with a part of
the K-M branch. A surface state of Ir(111), touching the renormalized
part of the $\pi^{*}$ band is also visible. It influences the analysis
of the MDC peak positions and widths in a small range of energies,
between -25~meV and -70~meV. These data will have to be interpolated.

\vspace{6mm}

\includegraphics[scale=0.8]{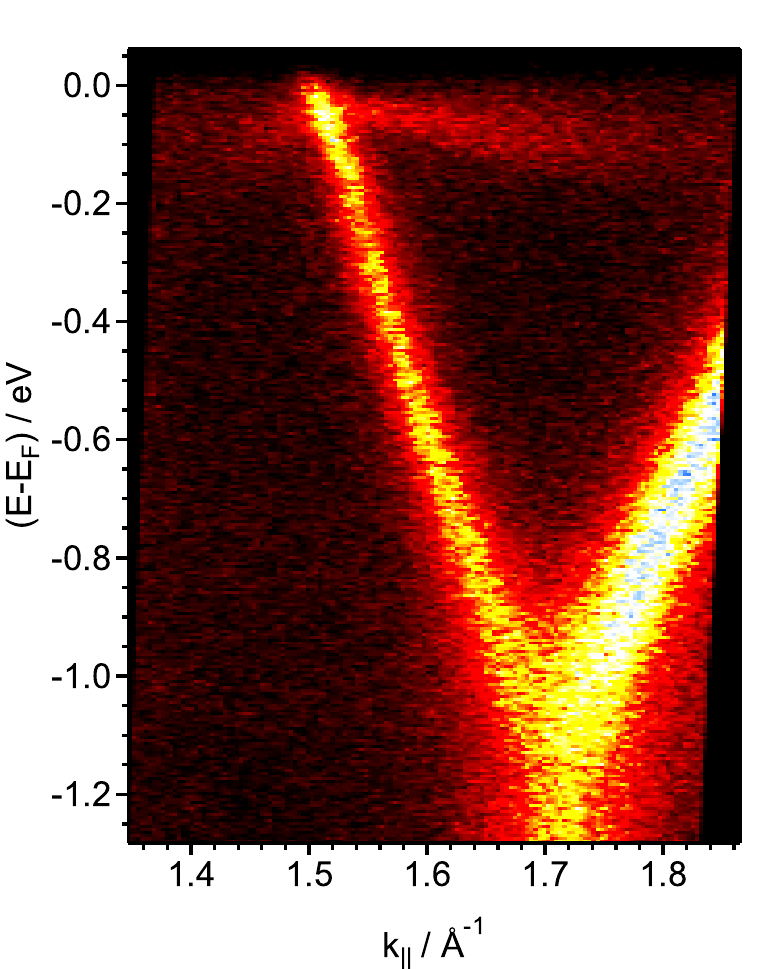}

\vspace{6mm}

We now show the data on FWHM and the position of MDC peaks extracted
from the above spectrum, along the K-$\Gamma$ branch of the $\pi^{*}$
band. The widths are comparable to those on the K-M branch, but the
noise in the data is obviously larger:

\vspace{6mm}

\includegraphics[width=7cm]{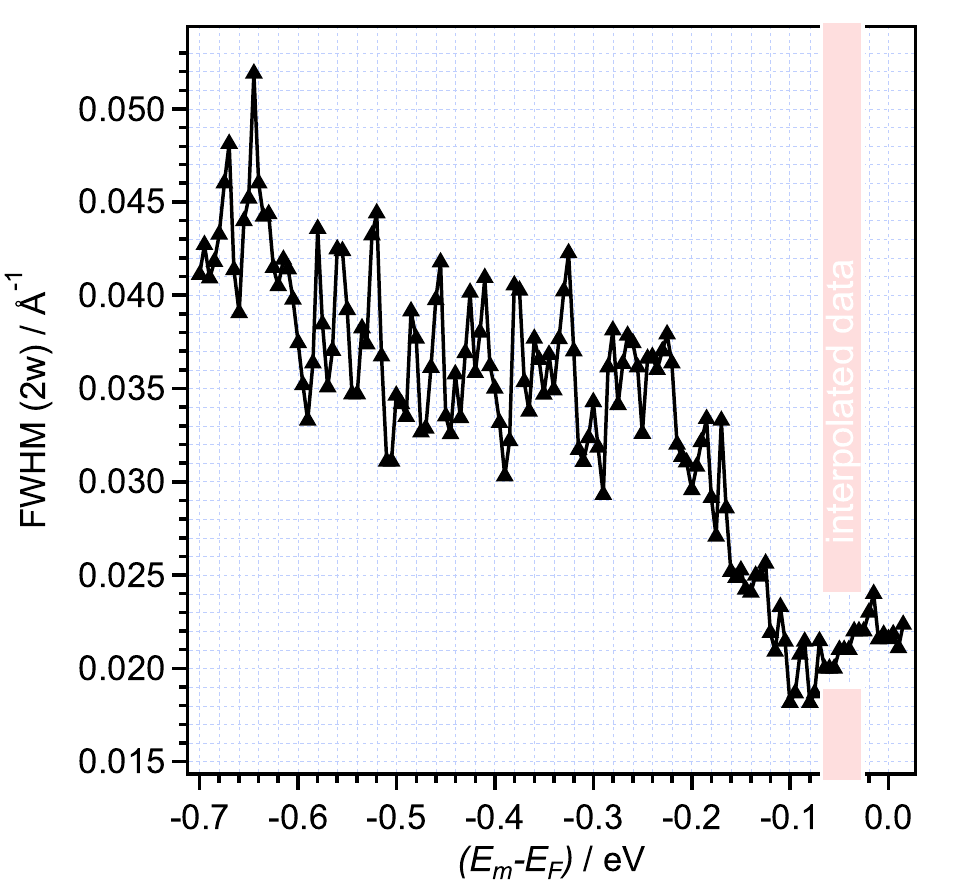}\includegraphics[width=8cm]{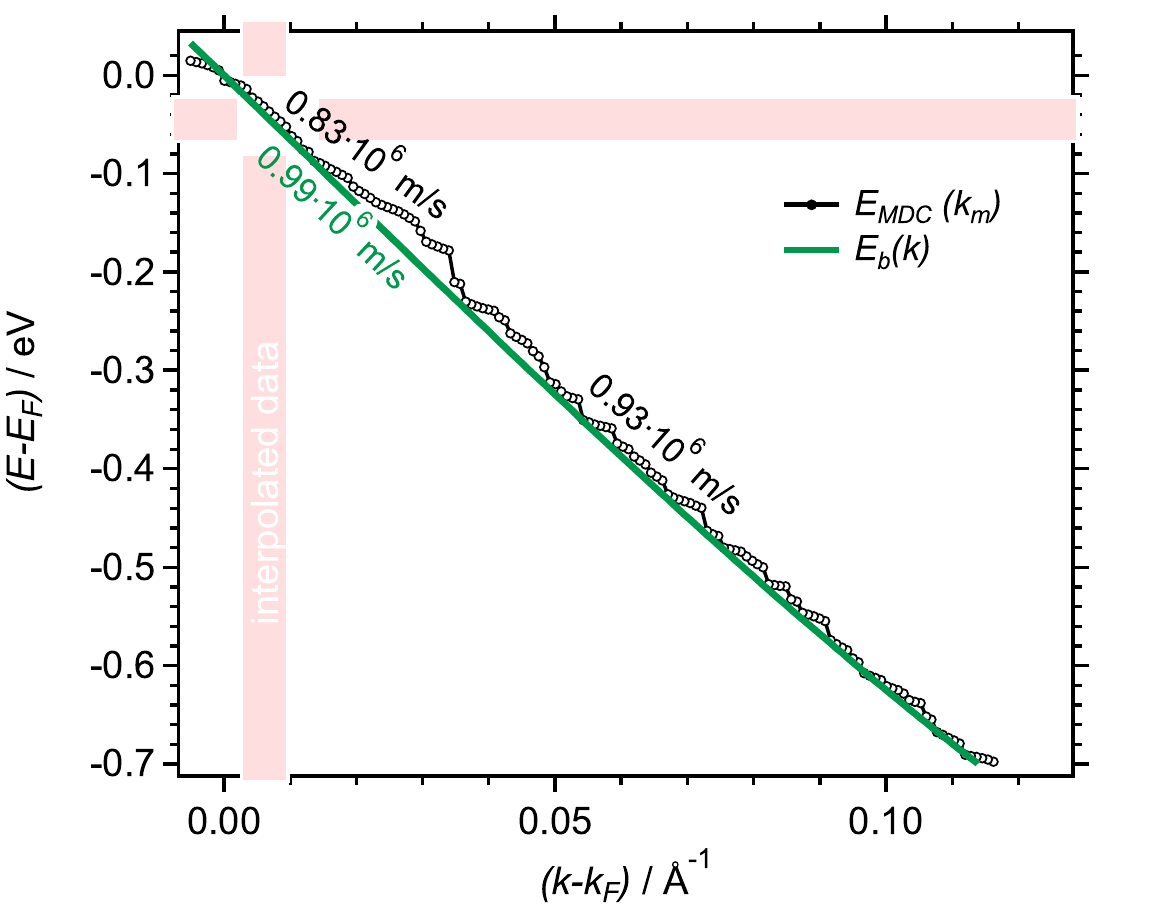}

\vspace{6mm}

Applying our self-consistent recursive procedure to the data from
the K-$\Gamma$ branch, we obtain the bare band (shown above, along
with the dispersion of the renormalized band), and the parts of the
self energy, $\Sigma=\Re\Sigma+i\,\Im\Sigma$:

\vspace{6mm}

\includegraphics[scale=0.8]{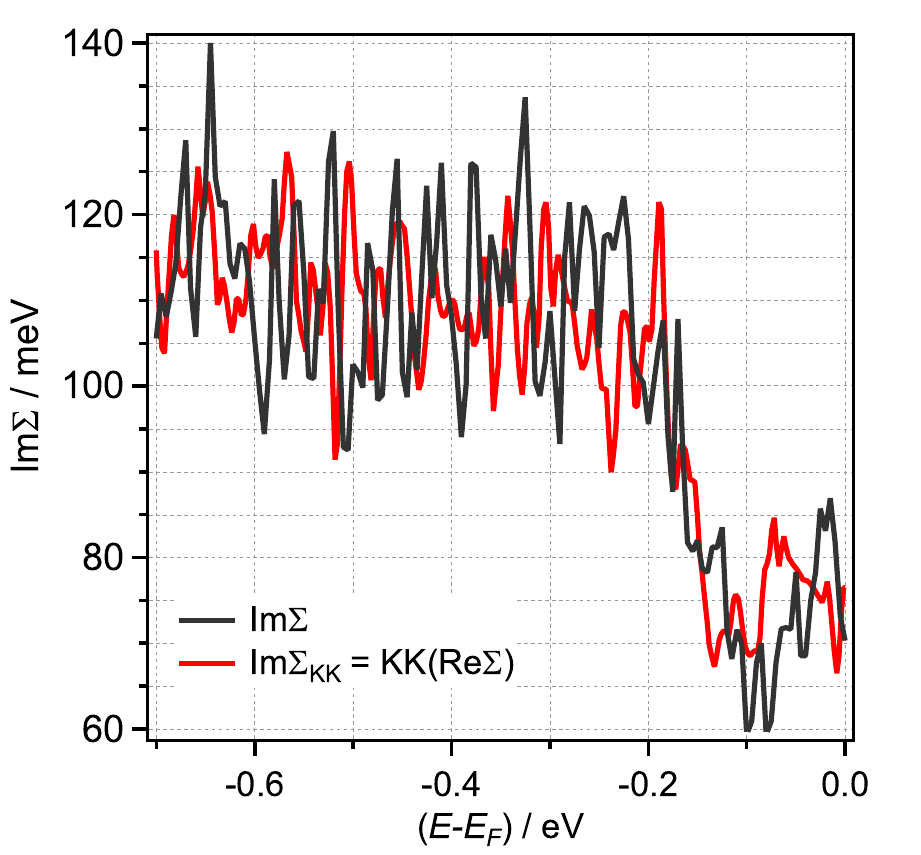}\includegraphics[scale=0.8]{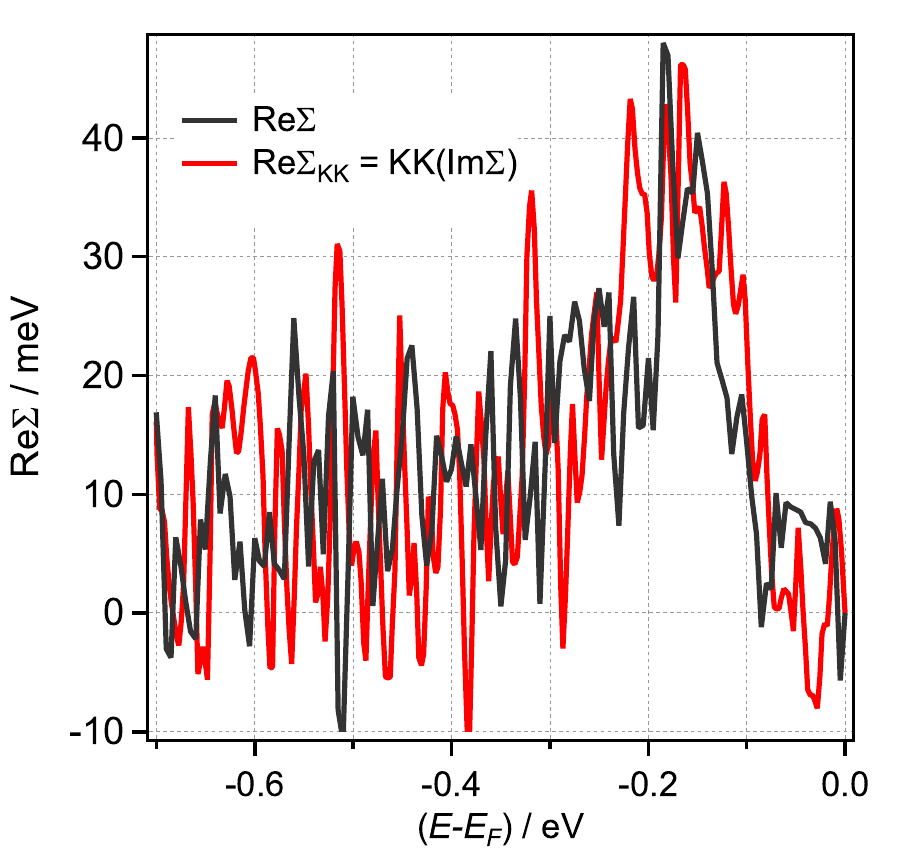}

\newpage{}
\begin{itemize}
\item \emph{Can you simulate a spectral function with the same lambda ($\lambda=0.2$)
in the K-M and K-$\Gamma$ direction. It would be interesting to see
if the different dispersion in the two directions is responsible for
the kink being more obvious in the KM direction.}
\end{itemize}
We simulate this by first recognizing that the same e-ph coupling
parameter $\lambda=0.2$ for the phonon of energy $\omega_{ph}=170$~meV
means, by equation (6) of our article, the same step in the imaginary
part of the self energy. This would be $\Delta\Im\Sigma=55\,\mathrm{meV}$
for both branches. The real part of the self energy is obtained by
the Kramers-Kronig transformation of the imaginary part, assuming
particle-hole symmetry:

\vspace{4mm}

\includegraphics[scale=0.7]{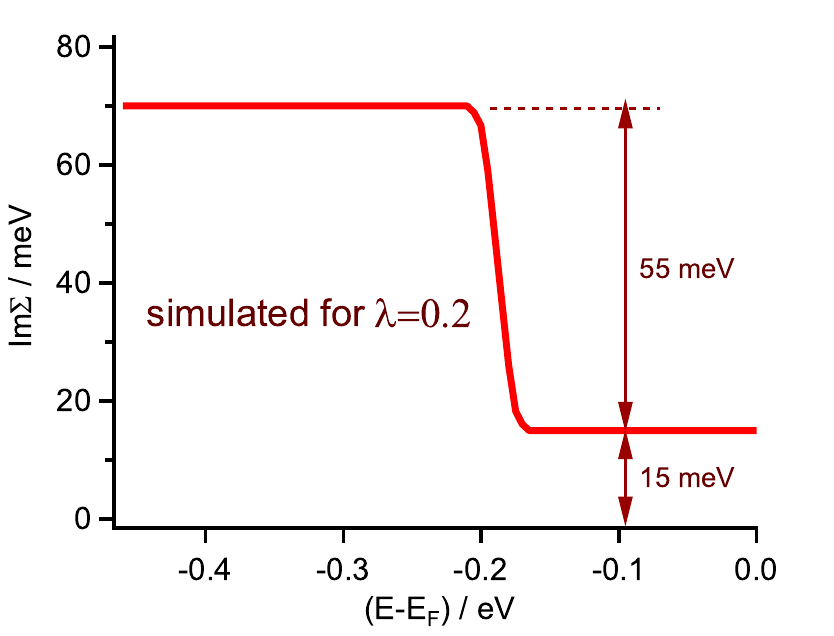}\includegraphics[scale=0.7]{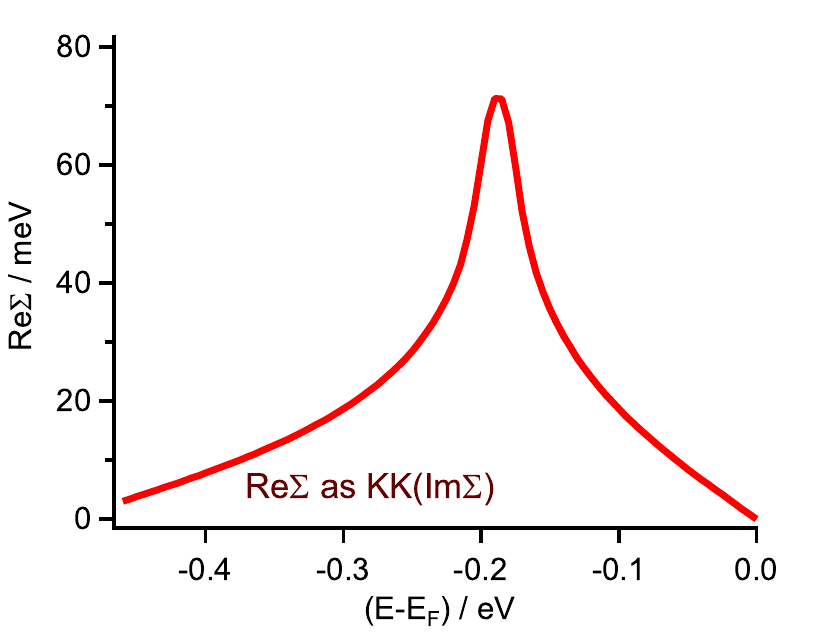}

\vspace{4mm}

We then take two linear bare bands, one mimicking the dispersion of
the K-$\Gamma$ branch (band velocity $1\cdot10^{6}$~m/s), the other
the dispersion of the K-M branch (band velocity $0.6\cdot10^{6}$~m/s),
and simulate the spectral function by equation (1) from our article:

\vspace{6mm}

\includegraphics[scale=0.9]{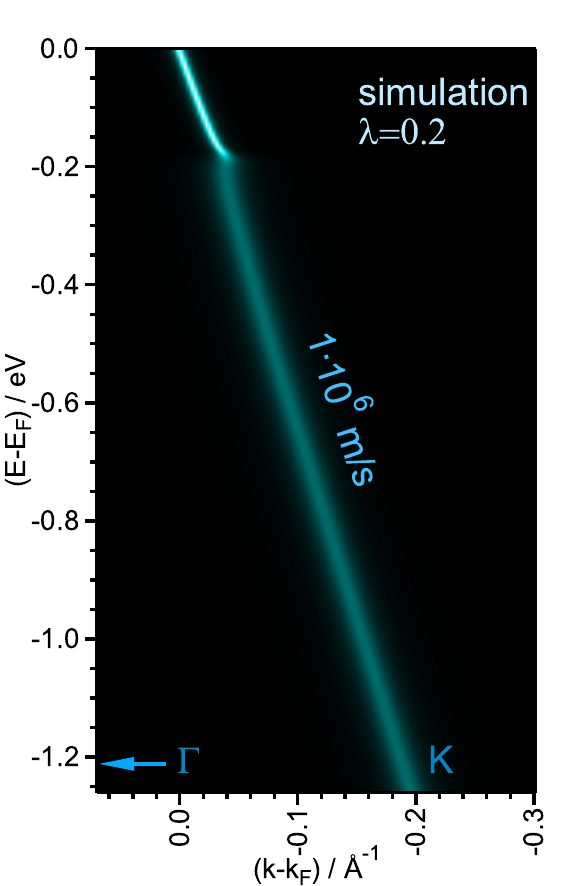}\qquad{}\includegraphics[scale=0.9]{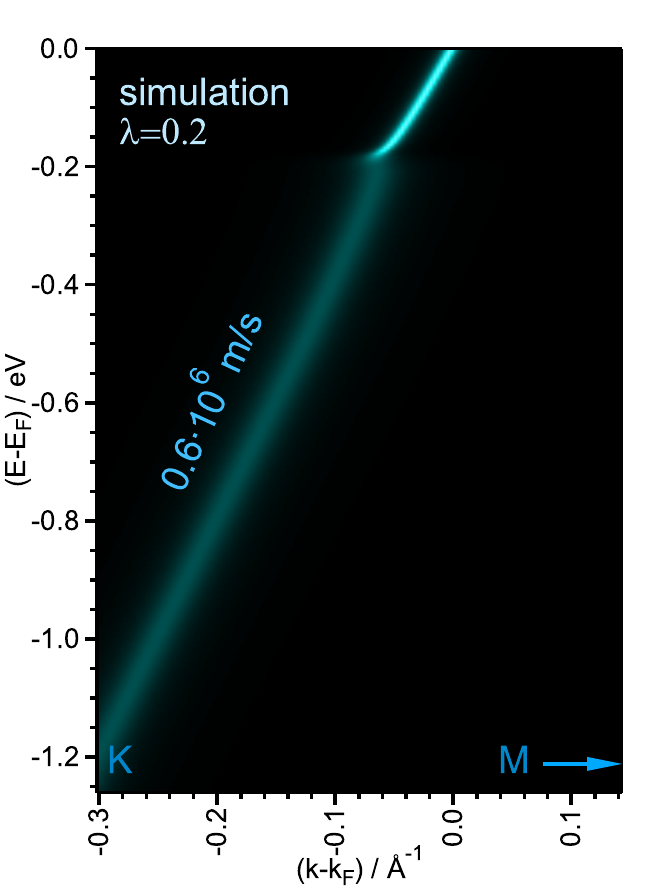}

The renormalized bands can be either extracted from the spectral function
following the dispersion of MDC peaks, or calculated by equation (2).
This is done self-consistently, starting with $E_{m}^{0}=E_{b}(k)+\Re\Sigma(E_{b}(k))$
and iterating over $E_{m}^{i+1}=E_{b}(k)+\Re\Sigma(E_{m}^{i})$. The
result can be immediately compared to the experimental data. Even
the simple linear band closely follows the measured band in a wide
range below the kink, as well as its renormalized part follows the
band above the kink. And indeed, the kink is more obvious for the
slowly dispersing band, i.e. the band simulating the dispersion along
the K-M direction. 

\vspace{6mm}

\includegraphics[scale=0.9]{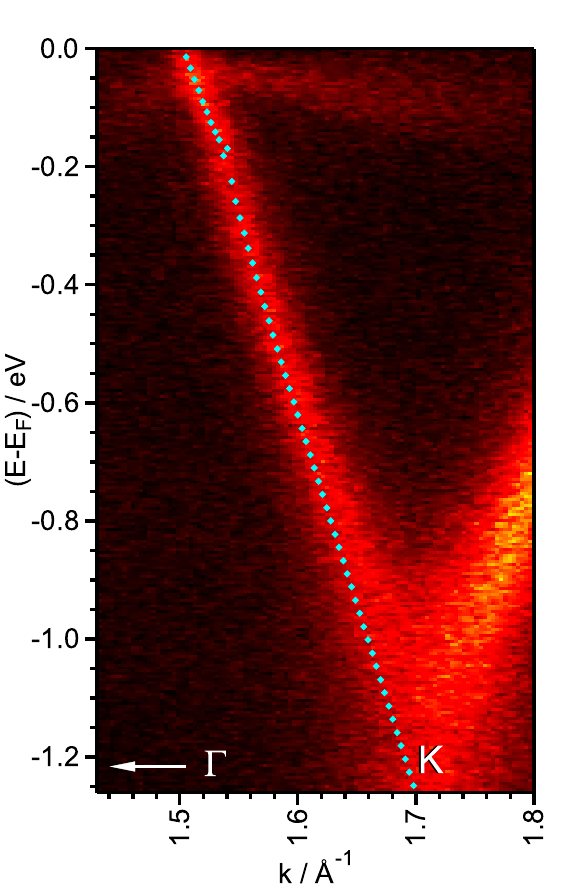}\qquad{}\includegraphics[scale=0.9]{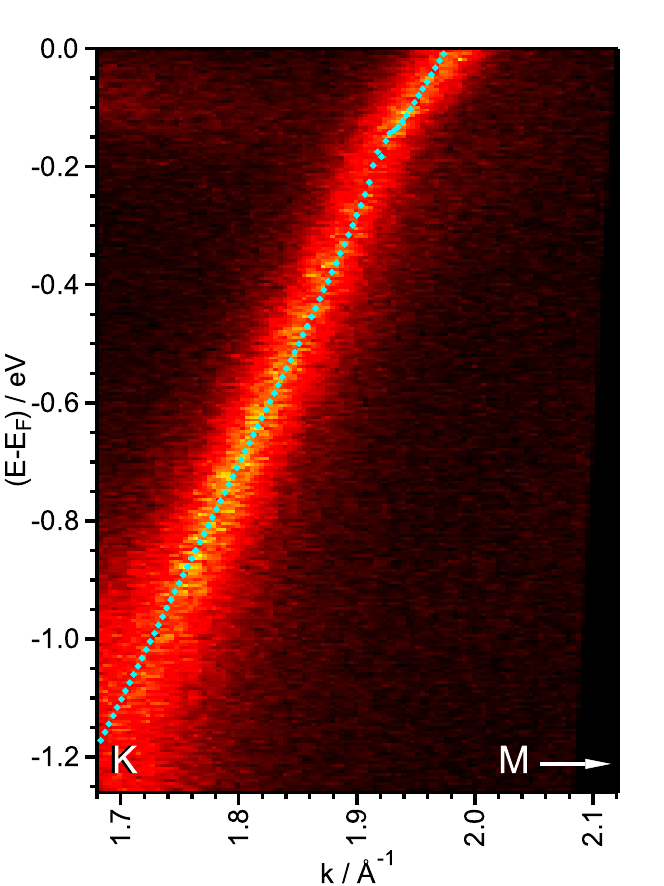}
}\end{spacing}

\end{document}